\documentclass[twocolumn,trackchanges, twocolappendix]{aastex7}

\usepackage{amsmath}
\usepackage{physics}
\usepackage{bm}
\usepackage{multirow}
\usepackage{textcomp}
\usepackage{mathcomp}
\usepackage{soul}

\usepackage{color}
\hypersetup{colorlinks=true, linkcolor=black, citecolor=blue,}

\begin{document}

\title{Investigation of Solar Wind Speed Characteristics Using IPS Observations and the PFSS+SCS Model}

\author[orcid=0009-0007-8101-9668, gname=Kyogo, sname=Tokoro]{Kyogo Tokoro} 
\affiliation{The University of Tokyo, Department of Earth and Planetary Science}
\email{t.kyogo1056.0706@eps.s.u-tokyo.ac.jp}

\author[orcid=0000-0002-7136-8190, gname=Munehito, sname=Shoda]{Munehito Shoda} 
\affiliation{The University of Tokyo, Department of Earth and Planetary Science}
\email{shoda.m.astroph@gmail.com}

\author[orcid=0000-0001-7891-3916, gname=Shinsuke, sname=Imada]{Shinsuke Imada} 
\affiliation{The University of Tokyo, Department of Earth and Planetary Science}
\email{imada@eps.s.u-tokyo.ac.jp}

\begin{abstract}
Understanding the relationship between solar wind speed and global coronal magnetic field is essential for space-weather forecasting and provides key diagnostics of the underlying acceleration mechanisms. Most observational studies to date have relied on near-ecliptic measurements dominated by slow wind, and the full range of wind speed, including fast wind, is insufficiently explored. Interplanetary scintillation (IPS) observations offer global coverage of wind speed; however, previous IPS-based studies have relied solely on the potential field source surface (PFSS) model, which does not adequately reproduce key features of the heliospheric magnetic field, potentially leading to the poor connectivity between the solar wind and its coronal source regions. Here we perform a comprehensive statistical analysis of the wind speed using IPS observations combined with PFSS and the Schatten current sheet (SCS) model. We find that the parameter $f_{\rm SS}/B_\odot$ bifurcates the solar wind into two distinct groups: one showing a strong negative correlation and the other showing no correlation. This grouping is better organized by the footpoint magnetic field strength and the distance from coronal hole boundary (DCHB) than by solar magnetic activity, suggesting that the two groups may reflect fundamentally different acceleration mechanisms.

\end{abstract}

\section{Introduction} \label{sec:intro}

Identifying the physical mechanisms that accelerate the solar wind and shape its velocity distribution remains one of the most important challenges in solar physics \citep{Parker_1958_ApJ, Cranmer_2002_SSR, Hansteen_2012_SSR}. Spatial and temporal variations in wind speed influence the large-scale structure of the heliosphere and a variety of space-weather phenomena \citep{Riley_2007_JASTP, Vrsnak_2007_A&A}. Therefore, elucidating the wind acceleration processes is crucial for advancing fundamental physics as well as for enabling practical applications. 

Multiple physical mechanisms are thought to contribute to wind acceleration, one of the leading candidates being the transport of energy and heating by Alfv\'en waves propagating upward from the lower solar atmosphere \citep{DePontieu_2007_Sci, McIntosh_2011_Natur, Morton_2025_ApJ}. Alfv\'en waves accelerate the solar wind through two primary pathways: plasma heating via nonlinear wave processes \citep{Suzuki_2005_ApJ, Usmanov_2014_ApJ, Magyar_2017_NatSR, Shoda_2019_ApJ, Reville_2020_ApJS}, which enhances gas pressure gradients, and direct momentum deposition via wave pressure \citep{Alazraki_1971_AandA, Hollweg_1973_JGR, Jacques_1977_ApJ, Rivera_2024_Science, Rivera_2025_ApJ}.

Another proposed mechanism for the solar wind acceleration is interchange reconnection \citep{Fisk_1999_JGR, Fisk_2003_JGR, Owens_2020_SoPh, Bale_2023_Natur, Shoda_2023_ApJ}. This process occurs between closed and open magnetic field lines, allowing plasma confined in closed loops and energy released by reconnection to be efficiently transferred to open flux tubes \citep{Wang_2004_ApJ, Rappazzo_2012_ApJ}. In particular, when interchange reconnection is active near coronal-hole boundaries, it can provide a significant source of mass and energy to the solar wind \citep{Iijima_2023_ApJ, Chen_2025_AandA}.

In the context of these acceleration mechanisms, several characteristic parameters have been identified as empirical signatures of solar wind speed. The super-radial expansion factor of magnetic flux tubes measured at the source surface, $f_{\rm SS}$, is one of the most widely used indicators, exhibiting an inverse correlation with the observed wind speed \citep{Wang_1990_ApJ, Arge_2000_JGR}. Because it links the geometrical properties of the source region to the efficiency of acceleration, $f_{\rm SS}$ is frequently incorporated into space-weather forecasting models \citep[e.g.,][]{Shiota_2014_SW, Shiota_2016_SW}. The distance from the coronal-hole boundary \citep[DCHB;][]{Riley_2001_JGR, Arge_2003_AIP} is another parameter commonly used to characterize the solar-wind source and is known to correlate with wind speed \citep[e.g.,][]{Owens_2008_SW}. Moreover, parameters that combine the expansion factor with the footpoint magnetic-field strength, such as $f_{\rm SS}/B_\odot$, have attracted attention as potentially more physically meaningful quantities \citep{Suzuki_2006_ApJ, Fujiki_2015_SoPh, Tokoro_2026_ApJ}.

A potential limitation of previous studies on solar wind speed is that they have been based primarily on observations near the ecliptic plane, where the slow wind dominates \citep[e.g.,][]{Wang_1990_ApJ, Riley_2015_SW, Wang_2020_ApJ}. As a consequence, the statistical characterization of the fast wind has remained insufficient. Only a few approaches have provided access to solar wind measurements at high heliographic latitudes, most notably the Ulysses mission \citep{Smith_1995_GRL, McComas_2008_GRL} and interplanetary scintillation (IPS) observations, although recent missions such as Solar Orbiter \citep{Muller_2020_A&A} and PUNCH \citep{Deforest_2022_aero} are expected to contribute significantly to high-latitude measurements. In particular, IPS observation is unique in that it has provided long-term, rotation-by-rotation global maps of the solar-wind speed, enabling statistical analyses of the wind properties over both low and high latitudes.

Observational studies using IPS data include \citet{Fujiki_2015_SoPh} and \citet{Tokumaru_2024_SoPh_DCHB}, although both rely solely on the potential field source surface (PFSS) model as the coronal magnetic-field extrapolation. The PFSS model \citep{Altschuler_1969_SoPh, Schatten_1969_SoPh} has been shown to reproduce the large-scale three-dimensional coronal magnetic field reasonably well despite its low computational cost \citep{Riley_2008_ApJ, Reville_2020_ApJS, Huang_2024_ApJL}. However, it is also known to exhibit shortcomings, including inaccuracies in streamer geometry, the location of coronal-hole boundaries, and the latitudinal uniformity of the open flux \citep{Smith_1995_GRL}, as reported by several studies \citep{Schatten_1971_CosEl, Riley_2008_ApJ}. Schatten current sheet (SCS) model \citep{Schatten_1971_CosEl, Schatten_1972_NASSP} was proposed as an improvement to address these issues, and may therefore be particularly important for IPS-based studies that seek to analyze global wind speed distributions.

In this study, we combine IPS observations with the PFSS+SCS model to re-examine the relationship between the wind speed and various flux-tube parameters. This approach allows us to investigate the characteristics of the wind speed and to discuss the underlying acceleration mechanisms.

\section{Method}

\subsection{IPS Observations}
In this work, we use the solar wind speed data derived from observations of IPS, a radio scattering phenomenon caused by electron density irregularities in the solar wind \citep{Kojima_1990_SSR, Asai_1995_JGG, Tokumaru_2011_RaSc, Tokumaru_2013_PJAB}. The IPS observations have been conducted at the Institute for Space-Earth Environmental Research of Nagoya University using three or four radio telescopes at 327 MHz. The solar wind velocity is derived by the cross-correlations among the telescopes. The wind velocity distribution is backmapped onto a reference sphere using computer-assisted tomography (CAT) method. The spatial resolution of the IPS data on the synoptic map is approximately $15 \tcdegree$.

\subsection{PFSS+SCS Model}

In this study, PFSS model in the vicinity of the Sun and SCS model in the outer region are used as coronal magnetic field extrapolation methods. An advantage of employing the SCS model in addition to the PFSS model is that it allows the latitudinal uniformity of the radial magnetic field to be reproduced beyond the source surface, which cannot be achieved with the PFSS model alone \citep{Smith_1995_GRL}. Because the SCS model incorporates flux-tube expansion above the source surface, the magnetic connectivity between the solar surface and the solar wind is altered, making it meaningful to re-examine the solar wind speed statistics using this combined model. In this study, following \citet{McGregor_2008_JGR} and \citet{Meadors_2020_SW}, we connect the PFSS and SCS models through an interface region. Specifically, after computing the PFSS solution from the solar surface to the source surface, the SCS model is constructed by setting a radius slightly below the source surface as its inner boundary, where the PFSS solution is applied as the boundary condition. Above the inner boundary of the SCS model, the magnetic field is described by the SCS solution. This treatment mitigates the magnetic-field discontinuities that otherwise arise between the PFSS source surface (where $B_\theta = B_\phi = 0$) and the inner boundary of the SCS model (where $B_\theta \ne 0$ and $B_\phi \ne 0$).

The PFSS+SCS framework contains three free parameters: the outer boundary of the PFSS model $r_{\rm PFSS,out}$, the value of $\varepsilon$ (i.e., the inner boundary of the SCS model $r_{\rm SCS,in}$), and the outer boundary of the SCS model $r_{\rm SCS,out}$. Although the widely used approach is to fix $r_{\rm PFSS,out}$ at $2.5 R_\odot$ \citep{Hoeksema_1983_JGR}, in this study, we adopt the method proposed by \citet{Shoda_2025_ApJ}, in which the source-surface height is determined from the open magnetic flux. The uncertainty from this method is discussed in Appendix \ref{app:pfssout}. To minimize the influence of the interface region on the open flux, we fix $\varepsilon = 0.1 R_\odot$. We then compute the average open flux at 1 au from the daily OMNI database for each CR and select, with an accuracy of $0.1 R_\odot$, the source-surface height that best matches the observed open flux. We note that because the entire SCS domain consists of open magnetic field lines, the open flux remains constant for $r \ge r_{\rm SCS,in}(= r_{\rm PFSS,out} - 0.1 R_\odot)$. The SCS outer boundary $r_{\rm SCS,out}$ has only a minor influence on the analysis (see Appendix \ref{app:scsout}). In this study, however, we set $r_{\rm SCS,out} = 10 R_\odot$ so that the radial magnetic field becomes sufficiently latitudinally uniform several solar radii beyond $r_{\rm PFSS,out} (\sim 2R_\odot)$, consistent with \citet{Shi_2024_ApJ}.

The specific design of the magnetic-field extrapolation model follows that described in \citet{Knizhnik_2024_FrASS}. The method utilizes a spherical harmonic expansion; the maximum spherical-harmonic degree $l_{\rm max}$ is set to 128 for the PFSS model, following \citet{Shiota_2014_SW}. In contrast, the SCS model uses a smaller value of \(l_{\rm max}=32\). This choice is justified because the SCS region lies outside the PFSS domain, and the contribution of high-order components, which decay rapidly with distance, becomes negligible at those larger radii.

\subsection{ADAPT Model}

As the photospheric boundary condition for the coronal field extrapolation, we use the synoptic maps of the radial magnetic field generated by the Air Force Data Assimilative Photospheric Flux Transport (ADAPT) model \citep{Worden_2000_SoPh,Arge_2010_AIPCS, Arge_2013_AIPC,Hickmann_2015_SoPh}. ADAPT model evolves the photospheric magnetic field on a full-Sun grid using a surface flux-transport model that includes differential rotation, meridional flow, and supergranular diffusion. For the period considered here, we employ the ADAPT-KPVT/VSM product, in which the assimilated observations are line-of-sight full-disk magnetograms from the NSO Kitt Peak Vacuum Telescope \citep[KPVT; 1970s--2003; ][]{Jones_1992_SoPh} and the SOLIS Vector Spectromagnetograph \citep[VSM; 2003; ][]{Keller_2003_ASPC, Henney_2006_ASPC}. The resulting ADAPT-KPVT/VSM maps provide a temporally consistent, assimilated time series of global radial-field distributions.

\subsection{Data Analysis}

We analyzed the solar wind speed and magnetic field data covering 32 Carrington Rotations (CRs) between 1992 and 2009. This period is chosen because, prior to 2009 when IPS observations were conducted by four telescopes, ADAPT-KPVT/VSM provides the only magnetic-field dataset that was generated consistently over a long time span and is suitable for comparative analysis. The CRs used for analysis are selected to minimize IPS data gaps and to ensure that each selected CR is separated by at least three CRs. The full list of analyzed CRs and their $r_{\rm PFSS,out}$ is provided in Table S1.

For each selected CR, we examine the correlation between the solar wind speed $v_{\rm IPS}$ and the parameters characterizing the magnetic flux-tube geometry derived from the PFSS+SCS model at a given latitude and Carrington longitude. Specifically, we focus on several characteristic parameters computed from the expansion factor on the source surface $f_{\rm SS}$ of each magnetic flux tube.
\begin{align}
    f_{\rm SS}(\theta,\phi) = \frac{B_{\rm SS}(\theta,\phi) r_{\rm SS}^2 }{ B_{\odot} R_\odot^2},
\end{align}
where $B_\odot$ is the magnetic field strength at $r = R_\odot$. In this study, the source surface is defined as $r=r_{\rm SCS,in}$. It should be noted that the IPS observations are projected onto $r = 2.5R_\odot$, whereas the characteristic parameters derived from the magnetic field extrapolation models are based on the flux-tube distribution at the outer boundary of the SCS model, $r_{\rm SCS,out} = 10R_\odot$.

The effective angular resolution of the synoptic solar wind velocity maps derived from IPS observations is approximately $15^\circ \times 15^\circ$ (Iwai 2025, private communication), which is significantly coarser than that of the ADAPT magnetic-field maps. To enable a fair comparison between the IPS-derived solar wind speeds and the magnetic-field-related parameters computed for individual magnetic flux tubes, it is therefore necessary to adjust the spatial resolution of the model-derived quantities. In this study, we apply a $15^\circ$ moving average to parameters defined along each magnetic flux tube. The data of open flux tubes are then resampled so that the sampling intervals remained uniform on the spherical surface. Specifically, to avoid latitudinal bias caused by the reduced effective longitudinal spacing toward the poles, the sampling interval in longitude $\Delta\phi$ at each latitude $\theta$ is adjusted such that $\Delta\phi = 15^\circ / \cos \theta$. 

\section{Result}



\begin{figure}[!t]
  \begin{center}
  \plotone{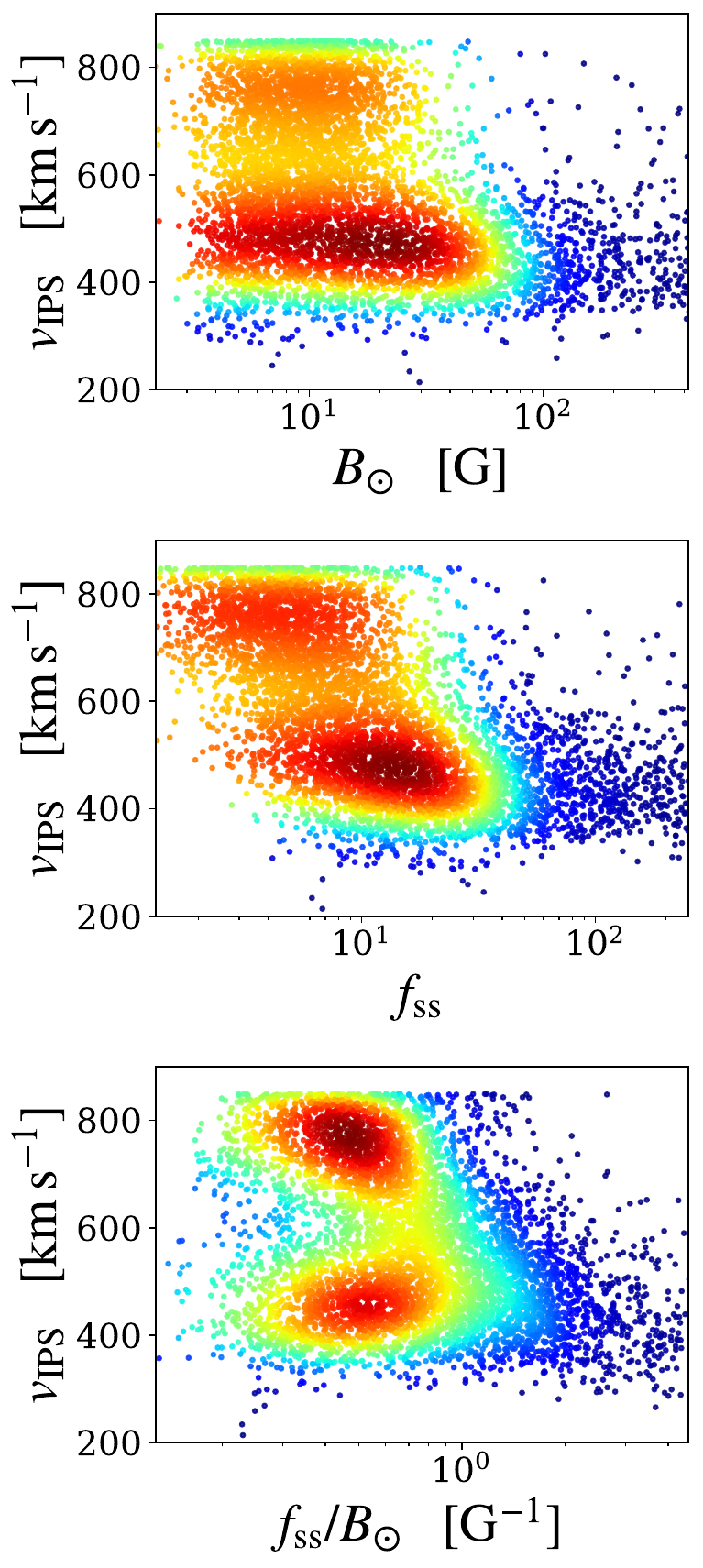}
  \caption{Relation between wind velocity by IPS observation $v_{\rm IPS}$ and each feature candidate. For all panels, the upper (lower) limit of the x-axis is set to twice (one half) the 95th (5th) percentile value, so as to facilitate visual interpretation of the correlation strength. The scatter plots are color-coded using Gaussian kernel density estimation computed using the SciPy scientific computing library \citep{Virtanen_2020_NatMe}, with high-density regions shown in red and low-density regions shown in blue. Top panel: the footpoint magnetic field strength, $B_\odot$. Middle panel: expansion factor at source surface $f_{\rm SS}$. Bottom panel: $f_{\rm SS}/B_\odot$.
  }
  \label{fig:feature and velocity}
  \end{center}
\end{figure}

Figure \ref{fig:feature and velocity} presents the correlations between the flux-tube parameters and wind speed for all open flux tubes within the analysis interval. In the top panel, the horizontal axis represents the footprint magnetic-field strength $B_\odot$. For strong footpoint magnetic fields of several tens of gauss or more, the resulting wind is predominantly slow. In contrast, for flux tubes with weaker footpoint fields, $B_\odot$ shows little correlation with the wind speed, producing both slow and fast flows. In the middle panel, the expansion factor at the source surface, $f_{\rm SS}$, exhibits an overall negative correlation with the wind speed. Yet for flux tubes with $f_{\rm SS}$ in the range of 5-10, the velocity distribution spans a wide range of 300-850 km s$^{-1}$, indicating that the correlation weakens for streams with moderate $f_{\rm SS}$ values. In contrast, the bottom panel shows that the scatter plot of the solar wind speed versus $f_{\rm SS}/B_\odot$, with point density color-coded, exhibits a clear two-branch structure corresponding to fast-wind and slow-wind populations.

\begin{figure}[!t]
  \begin{center}
  \plotone{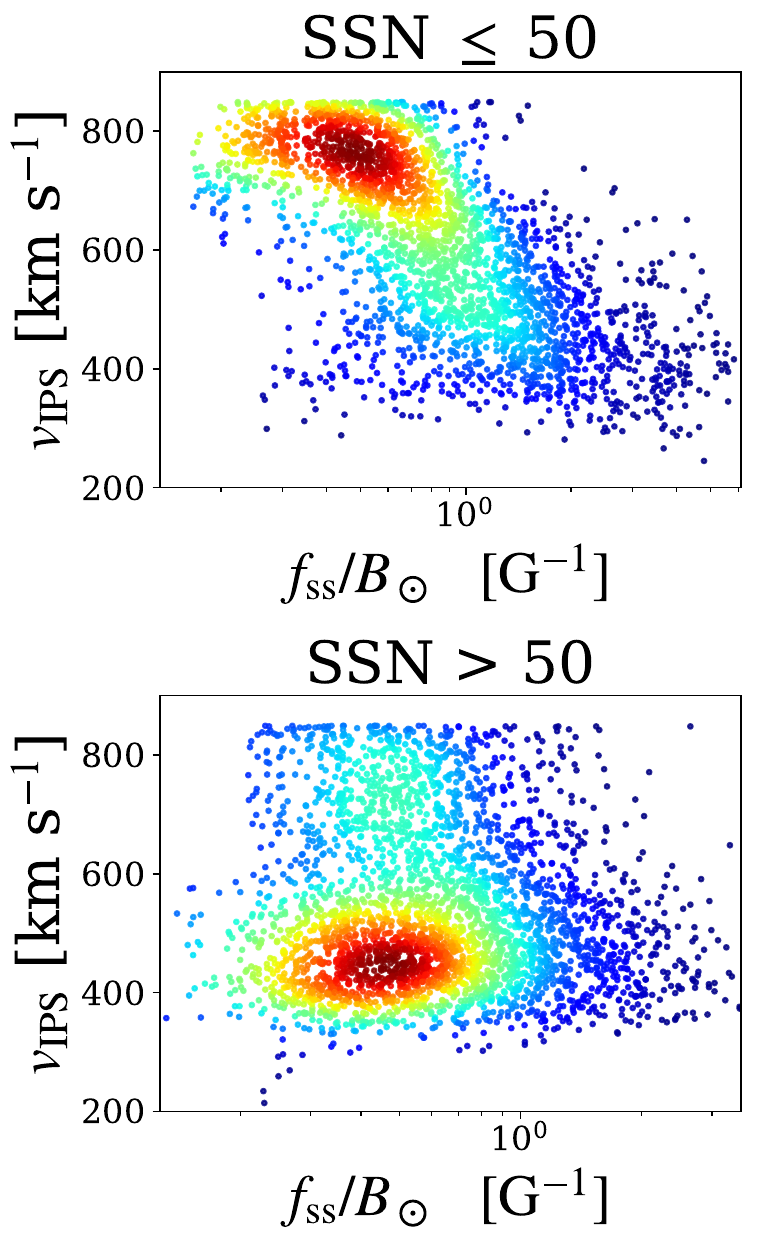}
  \caption{The format is the same as in Figure \ref{fig:feature and velocity}, but the horizontal axis is restricted to $f_{\rm SS}/B_\odot$. The upper panel presents the results for CRs with a sunspot number (SSN) of $\leq 50$ as a low-activity phase, whereas the lower panel shows those with SSN $> 50$ as a high-activity phase.
  }
  \label{fig: solar activity}
  \end{center}
\end{figure}

To investigate the origin of the two distinct branches seen in the bottom panel of Figure \ref{fig:feature and velocity}, we classify the magnetic flux tubes according to specific conditions and examine their correlation with the wind speed. Each CR is classified into the low-activity phase and the high-activity phase based on the sunspot number \citep[source: WDC-SILSO, Royal Observatory of Belgium, Brussels, https://doi.org/10.24414/qnza-ac80, ][]{SILSO_Sunspot_Number}. Specifically, we classify a CR as belonging to a high-activity phase when the 13-month smoothed sunspot number (SSN) is above 50, and to a low-activity phase otherwise. The corresponding correlations between $f_{\rm SS}/B_\odot$ and the wind speed $v_{\rm IPS}$ are shown in Figure \ref{fig: solar activity}. In the low-activity phase, the branch showing a strong negative correlation occupies most of the solar wind population, encompassing both fast and slow wind. In contrast, in the high-activity phase, the negative-correlation branch similar to that in the low-activity phase is still present, yet a population showing almost no correlation—primarily at $v_{\rm IPS} < 600 \: \mathrm{km \ s^{-1}}$ becomes dominant.

\begin{figure*}[!t]
  \begin{center}
  \plotone{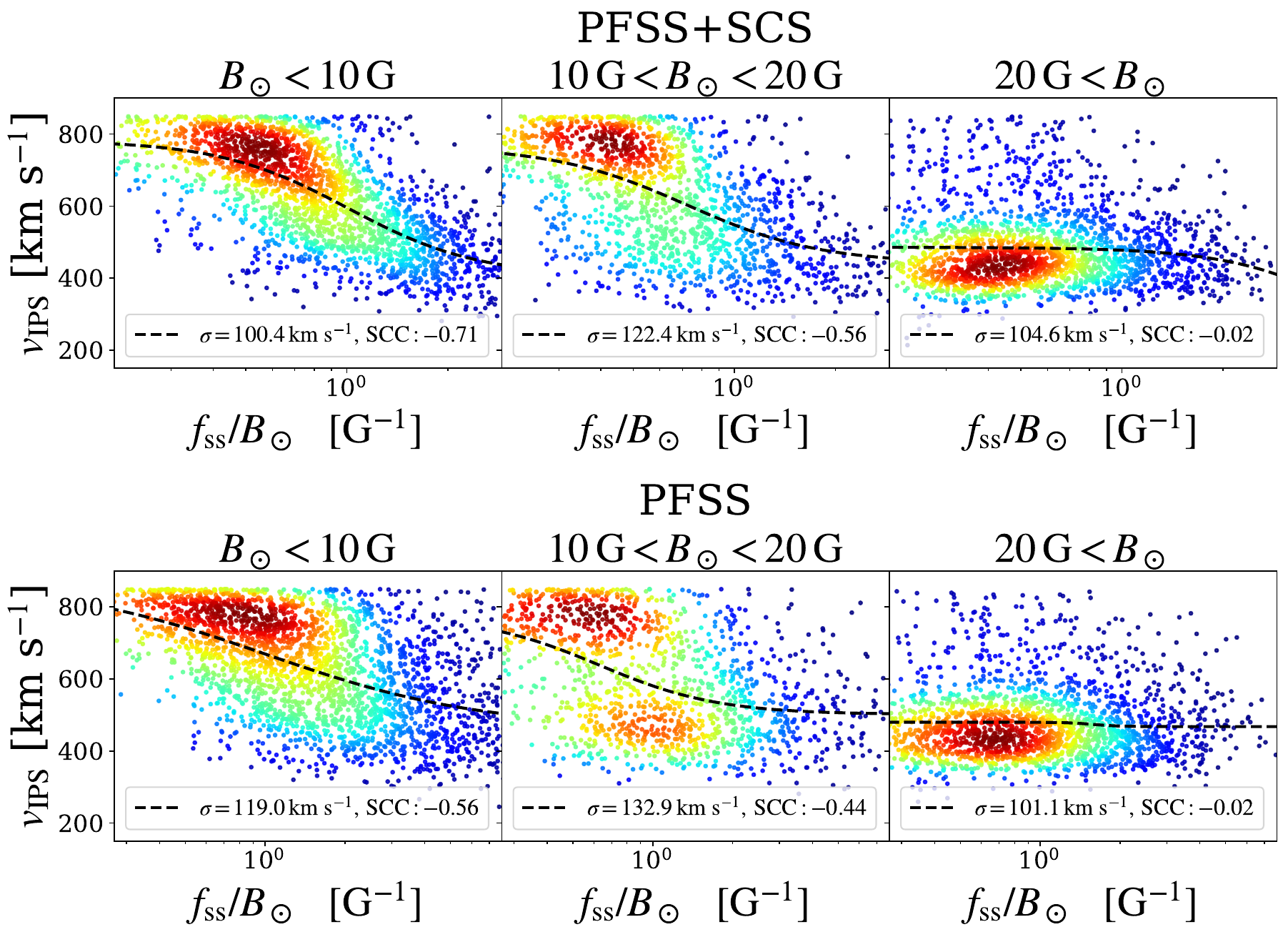}
  \caption{The format is the same as in Figure \ref{fig:feature and velocity}, but the horizontal axis is restricted to $f_{\rm SS}/B_\odot$. Each column is categorized by the value of the footpoint magnetic field $B_\odot$: left, $B_\odot < 10\ \rm{G}$; middle, $10\ \rm{G} < B_\odot < 20\ \rm{G}$; right, $20\ \rm{G} <B_\odot $. For reference, the best-fit  trend is shown as a black dashed line. The legend also reports the root mean square error of the fit, $\sigma$, and the Spearman correlation coefficient (SCC) between $f_{\rm SS}/B_\odot$ and $v_{\rm IPS}$. Each row corresponds to a different coronal magnetic field extrapolation model: the PFSS+SCS model (upper) and the PFSS model (lower).
  }
  \label{fig: foot point B}
  \end{center}
\end{figure*}

As an alternative way of classifying the data, we examine the correlation between $f_{\rm SS}/B_\odot$ and the wind speed by grouping all flux tubes according to the footpoint magnetic field strength $B_\odot$. The results are shown in the upper panels of Figure \ref{fig: foot point B}. For reference, we provide, for each group, the root-mean-squared error (RMSE) $\sigma$ of the best fit (black dashed line), where the fitting function is given as follows 
\begin{equation}
    v_{\rm fit}=a\tanh \bigg[ b \log \big( x/c \big) \bigg] + d.
\end{equation}
Furthermore, we also show the Spearman rank correlation coefficient (SCC) calculated using $f_{\rm SS}/B_\odot$ and the wind speed. For the weak-footpoint-field solar wind ($B_\odot < 10~ \mathrm{G}$; left panel), the wind speed spans a wide range of 400-850 km s$^{-1}$ and exhibits a relatively strong negative correlation with $f_{\rm SS}/B_\odot$ (SCC: $-0.71$). In contrast, for the strong-footpoint-field solar wind ($B_\odot > 20~ \mathrm{G}$; right panel), the speeds are confined to a narrower range of 350-550 km s$^{-1}$, and almost no correlation with $f_{\rm SS}/B_\odot$ is found (SCC: $-0.02$). The intermediate range ($10~ \mathrm{G} < B_\odot < 20~ \mathrm{G}$) shows the largest RMSE, which may reflect that this category shows characteristics belonging to both the weak‐ and strong-footpoint-field populations, thereby mixing different types of dependencies.

We additionally apply the same analysis to the case in which only the PFSS coronal magnetic field extrapolation model is used. The result is shown in the lower row of Figure \ref{fig: foot point B}. Comparing the PFSS+SCS cases with the PFSS-only cases reveals that the PFSS+SCS models produce a narrower distribution along the horizontal axis for the weak-footpoint-field group (upper left), resulting in a clearer contrast with the strong-footpoint-field group (upper right). In contrast, the PFSS-only models exhibit a broader horizontal spread for the weak-field group (lower left), making the distinction between the two populations separated by $B_\odot$ less evident. Consequently, for solar wind streams with $f_{\rm SS}/B_\odot \sim 1$-2, a large dispersion appears along the vertical axis.

\begin{figure*}[!t]
  \begin{center}
  \plotone{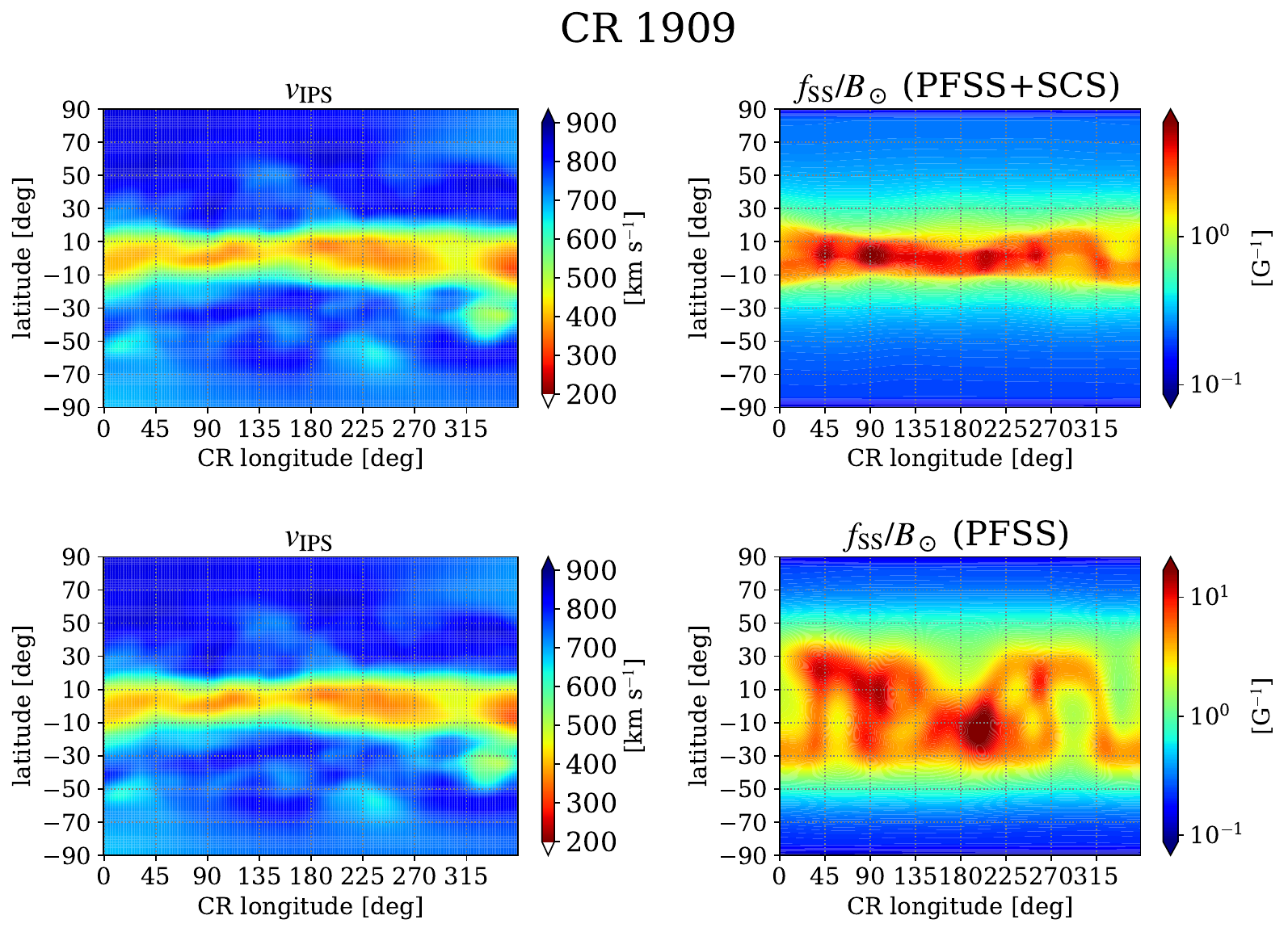}
  \caption{Synoptic maps for CR1909. The left column shows the solar wind speed derived from IPS observations ($v_{\rm IPS}$), which is identical in the upper and lower panels. The right column shows the distributions of $f_{\rm SS}/B_\odot$ obtained from the PFSS+SCS model (upper right) and the PFSS-only model (lower right) on the outer boundaty of the SCS model. For the right column, the color scale is set from half of the 5th percentile to twice the 95th percentile of the $f_{\rm SS}/B_\odot$ distribution.
  }
  \label{fig:scs CR1909}
  \end{center}
\end{figure*}

The difference in the horizontal distributions between the PFSS+SCS and PFSS-only models can be understood as follows. In the PFSS+SCS model, the magnetic field becomes more uniform at sufficiently large heliocentric distances. In contrast, when only the PFSS model is used, regions with small $B_{\rm SS}$—that is, regions with large $f_{\rm SS}/B_\odot$—tend to be overestimated in spatial extent. As a result, in the PFSS+SCS model, a fraction of fast wind streams that are originally connected to field lines with small $f_{\rm SS}/B_\odot$ are instead mapped to regions with large $f_{\rm SS}/B_\odot$. This leads to an apparent increase in the population of the fast wind associated with large $f_{\rm SS}/B_\odot$. As a specific example, Figure \ref{fig:scs CR1909} shows the distribution of the wind speed derived from the IPS observations and the distribution of $f_{\rm SS}/B_\odot$ on $r=r_{\rm SCS,out}$ derived from the coronal magnetic fields for CR1909. In this Carrington rotation, slow wind is distributed roughly between $-20^\circ$ and $+20^\circ$ latitude, while fast wind dominates at higher latitudes. In the PFSS+SCS model, regions with large $f_{\rm SS}/B_\odot$ are generally consistent with this structure. However, in the PFSS-only model, regions with large $f_{\rm SS}/B_\odot$ extend broadly from approximately $-40^\circ$ to $+40^\circ$ latitude, significantly overestimating their spatial extent.

\section{Summary and Discussion}

In this study, we have analyzed the wind speed characteristics over an 18-year period using the ADAPT-KPVT/VSM magnetograms, PFSS+SCS model, and IPS observations. The main findings can be summarized as follows. 
\begin{enumerate}
    \item Although the expansion factor at the source surface, $f_{\rm SS}$, shows an overall negative correlation with the wind speed, neither the footpoint magnetic-field strength $B_\odot$ nor the normalized parameter $f_{\rm SS}/B_\odot$ exhibits a simple dependence. 
    \item Instead, the wind speed distribution with respect to $f_{\rm SS}/B_\odot$ shows two distinct branches: one showing a strong negative correlation and another showing virtually no correlation. 
    \item The distinction between these two branches is determined more clearly by the magnitude of $B_\odot$ than by the magnetic activity level. For $B_\odot \gtrsim 20~ \mathrm{G}$, the solar wind tends to be slow regardless of the value of $f_{\rm SS}/B_\odot$, whereas for $B_\odot \lesssim 10~\mathrm{G}$, the wind speed can be either fast or slow depending on $f_{\rm SS}/B_\odot$. 
    \item Moreover, the contrast between the two dependence regimes is more pronounced when employing the PFSS+SCS model than when using the PFSS model alone.
\end{enumerate}

\subsection{Comparison with Previous Studies}

We first discuss the relationship between our results and previous observational studies. \citet{Fujiki_2015_SoPh} investigated the correlations among $B_\odot$, $f_{\rm SS}$, and $B_\odot/f_{\rm SS}$ (the reciprocal of $f_{\rm SS}/B_\odot$) using KPVT/VSM magnetograms, the PFSS model (with $r_{\rm PFSS,out}$ fixed at $2.5 R_\odot$), and IPS observations. Regarding the relationships of the wind speed with $B_\odot$ and $f_{\rm SS}$, this study and \citet{Fujiki_2015_SoPh} show similar tendencies. In both studies, a weak negative correlation is found between the solar wind velocity and $B_\odot$, and a weak but systematic dependence is found between the wind speed and $f_{\rm SS}$.

In contrast, our results differ from those of \citet{Fujiki_2015_SoPh} in the relationship between the wind velocity and $f_{\rm SS}/B_\odot$. \citet{Fujiki_2015_SoPh} reported that the solar-wind streams with $f_{\rm SS}<100$ and $B_\odot<5$ G depend on $B_\odot/f_{\rm SS}$, whereas those with $f_{\rm SS}>1000$ and $B_\odot>50$ G show no clear dependence. Since the relationship varies significantly from year to year, they concluded that no systematic relationship exists between the solar wind velocity and $B_\odot/f_{\rm SS}$. However, this study suggests that the differences in the dependence on $f_{\rm SS}/B_\odot$ arise from differences in the footpoint magnetic field strength. When classified by the footpoint magnetic field strength, systematic relationships between the solar wind velocity and$f_{\rm SS}/B_\odot$ can be identified. 

One possible cause for the different conclusions is the difference in the analysis approach. \citet{Fujiki_2015_SoPh} discussed the dependence only in terms of the mean value of each characteristic parameter within a given speed bin, and reported that the dependence of these mean values varied from year to year. In contrast, what we find is that examining the full distribution rather than mean values alone reveals a clearer dependence on the characteristic parameters. Moreover, as noted earlier, the SCS model enhances the dependence of the streams with weak $B_\odot$, suggesting that the year-to-year variability in the mean-value dependence may also be reduced.

We also compare our results with those of \citet{Wang_2020_ApJ}, who examined the correlations among $f_{\rm SS}$, $B_\odot$, $B_{\rm SS} (\propto B_\odot/f_{\rm SS})$, and the wind speed using magnetograms from the Mount Wilson Observatory and the Wilcox Solar Observatory, the PFSS model (with $r_{\rm PFSS,out}$ fixed at $2.5 R_\odot$), and 1-au observations retrieved from OMNI. Both \citet{Wang_2020_ApJ} and our work show a weak but systematic dependence of the solar wind speed on $f_{\rm SS}$. On the other hand, for the relationship between the wind speed and $B_\odot$, \citet{Wang_2020_ApJ} found almost no correlation, whereas this study finds a weak negative correlation. Furthermore, \citet{Wang_2020_ApJ} reported a correlation between the solar wind speed and $B_{\rm SS}$; however, they found that the nature of this dependence differs between solar maximum and solar minimum, limiting its usefulness as a universal characteristic parameter. In contrast, this study reaches a different conclusion, as described above.

\begin{figure}[!t]
  \begin{center}
  \plotone{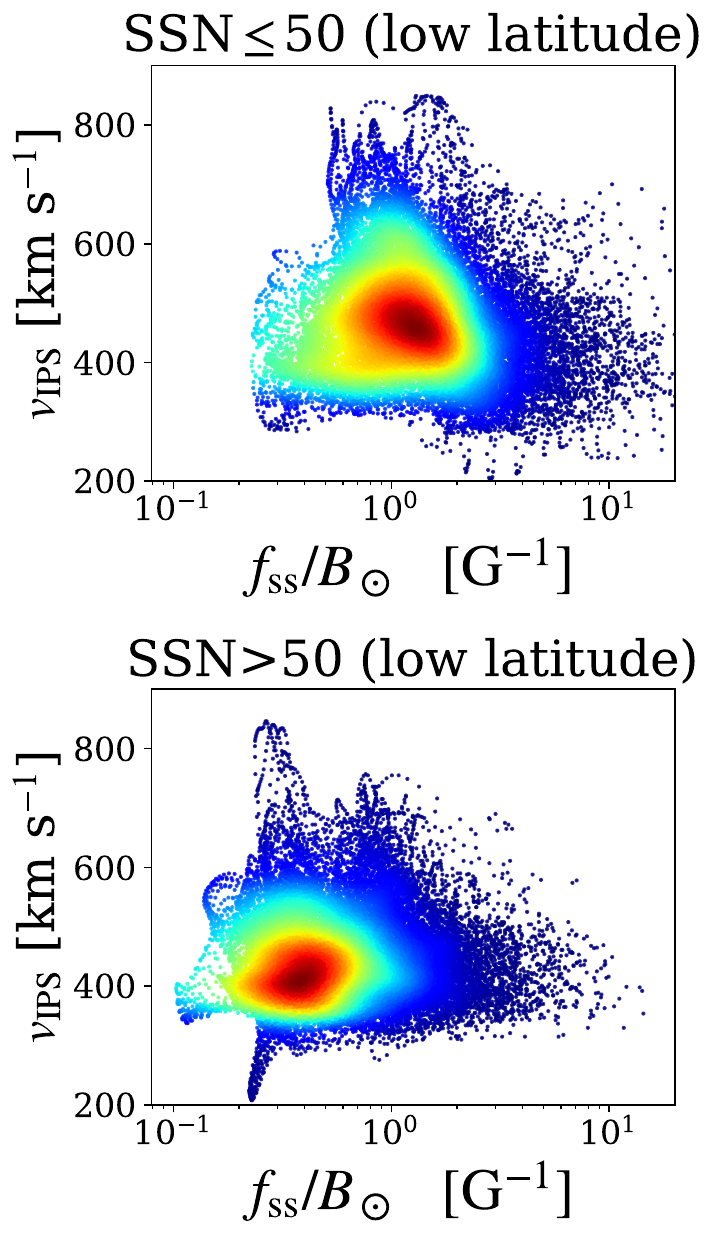}
  \caption{The format is the same as in Figure \ref{fig: solar activity}, but the solar wind data are restricted to the latitude range from $-20^\circ$ to $+20^\circ$, sampled at $1^\circ$ intervals, without applying any spatial averaging to the parameters; in addition, the horizontal axis ranges are identical between the upper and lower panels.
  }
  \label{fig:only ecliptic}
  \end{center}
\end{figure}

The discrepancy between \citet{Wang_2020_ApJ} and our work is attributable not only to the differences in the photospheric magnetograms and the field-extrapolation methods but also to the differences in the latitude ranges analyzed. IPS observations provide the global coverage of the wind speed, and during solar minimum in particular, they include more abundant measurements of fast wind data in high-latitude region in contrast to OMNI, which samples only near-Earth conditions. To assess the effect of sampling bias in latitude, we restrict the analysis to the low latitudes between $-20^\circ$ and $+20^\circ$, as shown in Figure \ref{fig:only ecliptic}. To ensure sufficient data coverage, the solar wind is sampled at $1^\circ \times 1^\circ$ resolution in latitude and longitude, and no spatial averaging is applied to $f_{\rm SS}/B_\odot$. In this restricted range, the values of $f_{\rm SS}/B_\odot$ during the low-activity phase (upper panel) are predominantly distributed above $\sim 1~\mathrm{G}^{-1}$, whereas during the high-activity phase (lower panel) they are concentrated around $\sim 0.4~\mathrm{G}^{-1}$. Thus, our analysis also indicates that the range of $B_{\rm SS}$ varies with solar activity. As discussed above, however, the variation in the dependence on wind speed does not originate from the solar-cycle phase itself, but rather from changes in the distribution of $B_\odot$.

\subsection{Possible Differences in Wind Acceleration Mechanisms}

As shown by \citet{Suzuki_2006_ApJ} and \citet{Tokoro_2026_ApJ}, the wind speed is expected to have a negative correlation with $f_{\rm SS}/B_\odot$ under wave-driven scenario. Taken together with our results, this suggests that the dependence of the wind speed on $f_{\rm SS}/B_\odot$ observed for weak footpoint magnetic fields can be explained within the framework of wave-driven models. In contrast, the absence of a dependence on $f_{\rm SS}/B_\odot$ in cases with strong footpoint magnetic fields implies that a different acceleration mechanism may be required to explain the observed behavior.

\begin{figure}[!t]
  \begin{center}
  \plotone{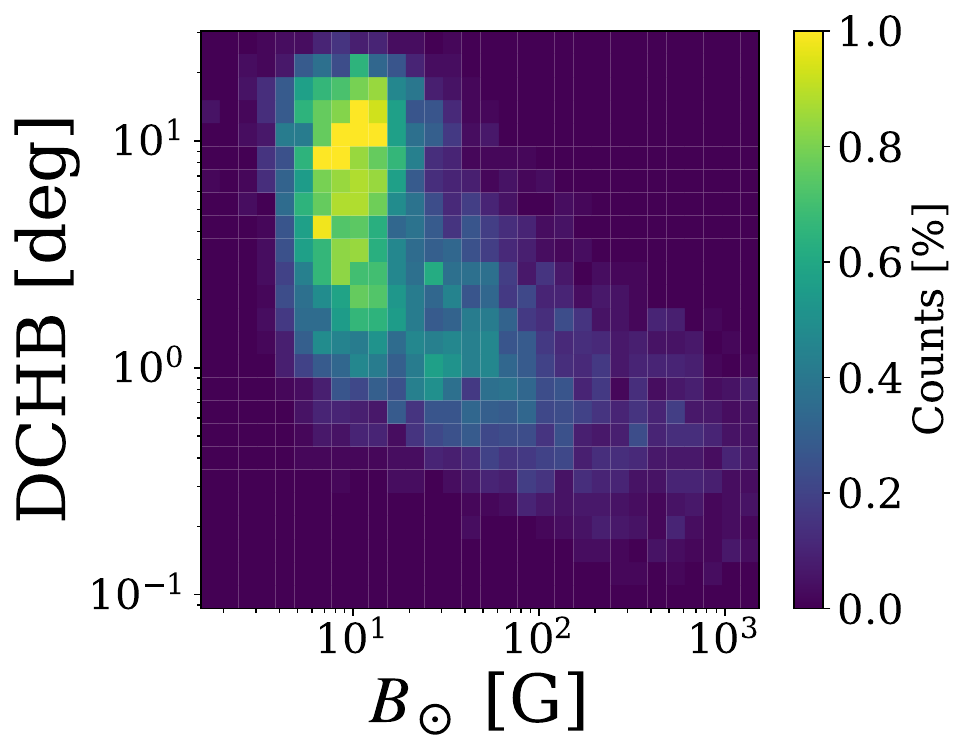}
  \caption{A two-dimensional histogram of $B_\odot$ and DCHB for all streams over the entire analysis period.
  }
  \label{fig: DCHB Bodot}
  \end{center}
\end{figure}

As a further analysis of the differences in dependence, we examine the relationship with the DCHB \citep{Riley_2001_JGR, Arge_2003_AIP}. Here, the coronal hole boundary (CHB) is identified using the $S$ factor \citep{Tokumaru_2024_SoPh_pseudo}, which is defined as follows.
\begin{equation}
    S=\sqrt{\left(\frac{\delta \varphi_\odot}{\delta \varphi_{\rm SCS,out}}\right)_{\rm lat}^2+\left(\frac{\delta \varphi_\odot}{\delta \varphi_{\rm SCS,out}}\right)_{\rm lon}^2},
\end{equation}
where $(\delta \varphi_\odot / \delta \varphi_{\rm SCS,out})_{\rm lat}$ and $(\delta \varphi_\odot / \delta \varphi_{\rm SCS,out})_{\rm lon}$ represent, respectively, how much the angular separation at $r=r_\odot$ differs from that at $r=r_{\rm SCS,out}$ for two adjacent flux tubes in the latitudinal and longitudinal directions at $r=r_{\rm SCS,out}$. A large value of the $S$ factor means that the footpoint coordinates of the open flux become discontinuous, thus indicating the presence of CHB. Specifically, the CHB is defined as the set of pixels satisfying $S_{\rm CHB} > \bar{S} + 3\sigma_S$, where $\bar{S}$ is the mean value of $S$ for each CR and $\sigma_S$ is the standard deviation. The DCHB for each pixel is computed as the minimum angular distance between that flux tube and the CHB on the $r = r_\odot$ surface.

Figure \ref{fig: DCHB Bodot} shows the two-dimensional histogram of $B_\odot$ and DCHB. As indicated, the dominant component of the open flux lies at DCHB values of several degrees or more and within the range $4~ \mathrm{G} < B_\odot < 20~ \mathrm{G}$. In contrast, for regions with $B_\odot > 20~ \mathrm{G}$, nearly all flux tubes have DCHB values smaller than 3 degrees, with virtually no population at large DCHB. This result suggests that the transition in the solar-wind properties around a footpoint field strength of 20 G found in our analysis corresponds to the change in the dependence of the DCHB on the footpoint field strength across the same threshold.

In summary, the dependence of the wind speed on $f_{\rm SS}/B_\odot$ in weak-footpoint-field regions supports a wave-driven solar wind. In contrast, the corresponding behavior in strong-footpoint-field regions—where DCHB is small—suggests that a different acceleration mechanism may operate, with reconnection/loop-opening processes as promising candidates.

\subsection{Limitation and Possible Applications}

Among the limitations of this study, the most significant is that the characteristic quantities for each flux tube were averaged over angular scales of approximately $15^\circ$ to match the spatial resolution of the IPS observations. While this coarse graining partially reduces the error associated with wind acceleration between the solar surface and $r = r_{\rm SCS,out} = 10R_\odot$ \citep{Riley_2015_SW}, it inevitably neglects finer spatial structures in the wind speed, such as those associated with pseudostreamers \citep{Wang_2007_ApJ, Riley_2012_SoPh, Wang_2012_ApJ, Riley_2015_SW}.

Another limitation of this study is the use of the PFSS+SCS model as the coronal magnetic field model. As discussed along Figures \ref{fig: foot point B} and \ref{fig:scs CR1909}, the PFSS+SCS model provides a significant improvement over the PFSS model in terms of the global connectivity of magnetic field lines. However, it can introduce artificial magnetic field structures, such as discontinuities in field-line connectivity at the interface between the PFSS and SCS domains, caused by the imposed current sheet \citep{McGregor_2008_JGR, Knizhnik_2024_FrASS}. Therefore, the use of more advanced models capable of reconstructing a more realistic coronal magnetic field configuration, such as the current sheet source surface model \citep{Zhao_1995_JGR} and outflow models \citep{Rice_2021_ApJ, Rice_2026_arXiv}, may enable further investigation of physically meaningful parameters.

Despite the limitations discussed above, this study suggests the dependence of the wind speed on $f_{\rm SS}/B_\odot$, as well as the need for more advanced models—such as the PFSS+SCS model—beyond the standard PFSS model in interpreting the IPS observations. Furthermore, a more detailed examination is necessary to elucidate the connection between wind acceleration mechanisms and their dependence on characteristic parameters such as $f_{\rm SS}/B_\odot$.

\begin{acknowledgments}
IPS observations were made under the solar wind program of the Institute for Space-Earth Environmental Research, Nagoya University. This work utilizes data produced collaboratively between Air Force Research Laboratory (AFRL) \& the National Solar Observatory (NSO). ADAPT \& SIFT model development is supported by AFRL. The input data utilized by ADAPT is obtained by NSO/NISP (NSO Integrated Synoptic Program). NSO is operated by the Association of Universities for Research in Astronomy (AURA), Inc., under a cooperative agreement with the National Science Foundation (NSF).

KT is supported by International Graduate Program for Excellence in Earth-Space Science (IGPEES), a World-leading Innovative Graduate Study (WINGS) Program, the University of Tokyo. MS and SI are supported by JSPS KAKENHI Grant Numbers JP24K00688,
JP25K00976, JP25K01052, by the grant of Joint Research by the National Institutes of Natural Sciences (NINS) (NINS program No. OML032402).
\end{acknowledgments}

\appendix

\section{Dependence on $r_{\rm SCS,out}$} \label{app:scsout}

\begin{figure*}[!t]
  \begin{center}
  \plotone{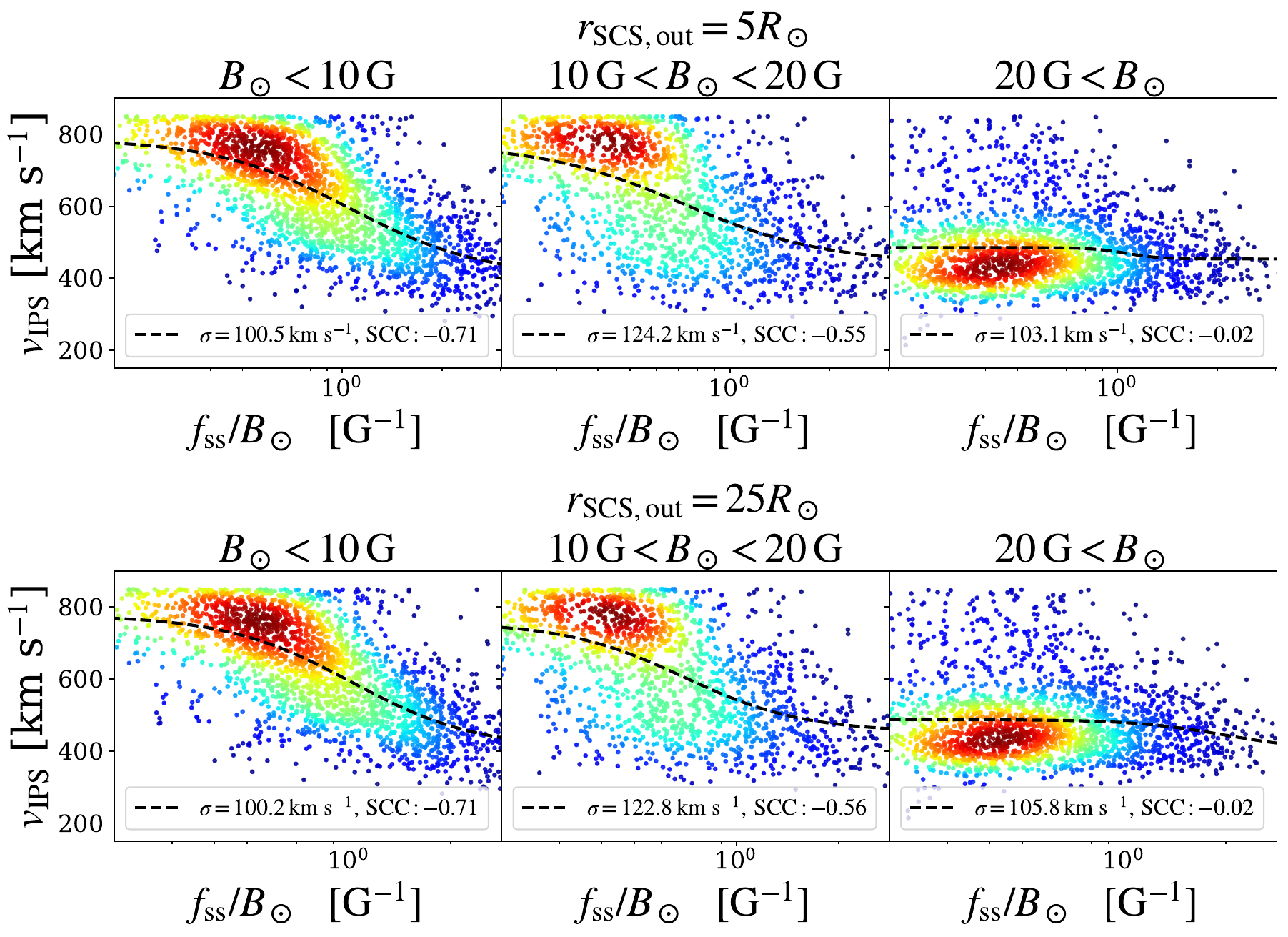}
  \caption{The same as the upper row of Figure \ref{fig: foot point B} but $r_{\rm SCS,out} = 5R_\odot$ (the upper row) and $r_{\rm SCS,out} = 5R_\odot$ (the lower row)
  }
  \label{fig:r_SCSout}
  \end{center}
\end{figure*}

In the main text, we adopt $r_{\rm SCS,out} = 10R_\odot$ following \citet{Shi_2024_ApJ}. Here, we verify that this assumption does not affect our conclusions. Figure \ref{fig:r_SCSout} shows the same analysis as the upper row of Figure \ref{fig: foot point B} in the main text: the upper row corresponds to the case with $r_{\rm SCS,out} = 5R_\odot$, while the lower row corresponds to $r_{\rm SCS,out} = 25R_\odot$. As is readily seen from the figure, although the positions of individual data points vary between the left and right columns, the overall trends remain unchanged.

For reference, the magnetic field lines shift by only a few degrees in latitude and longitude between $r = 5R_\odot$ and $r = 25R_\odot$, which is much smaller than the spatial resolution of approximately $15^\circ$ adopted in this study. Therefore, the choice of $r_{\rm SCS,out}$ does not significantly affect the results.

\section{Dependence on $r_{\rm PFSS,out}$} \label{app:pfssout}

\begin{figure*}[!t]
  \begin{center}
  \plotone{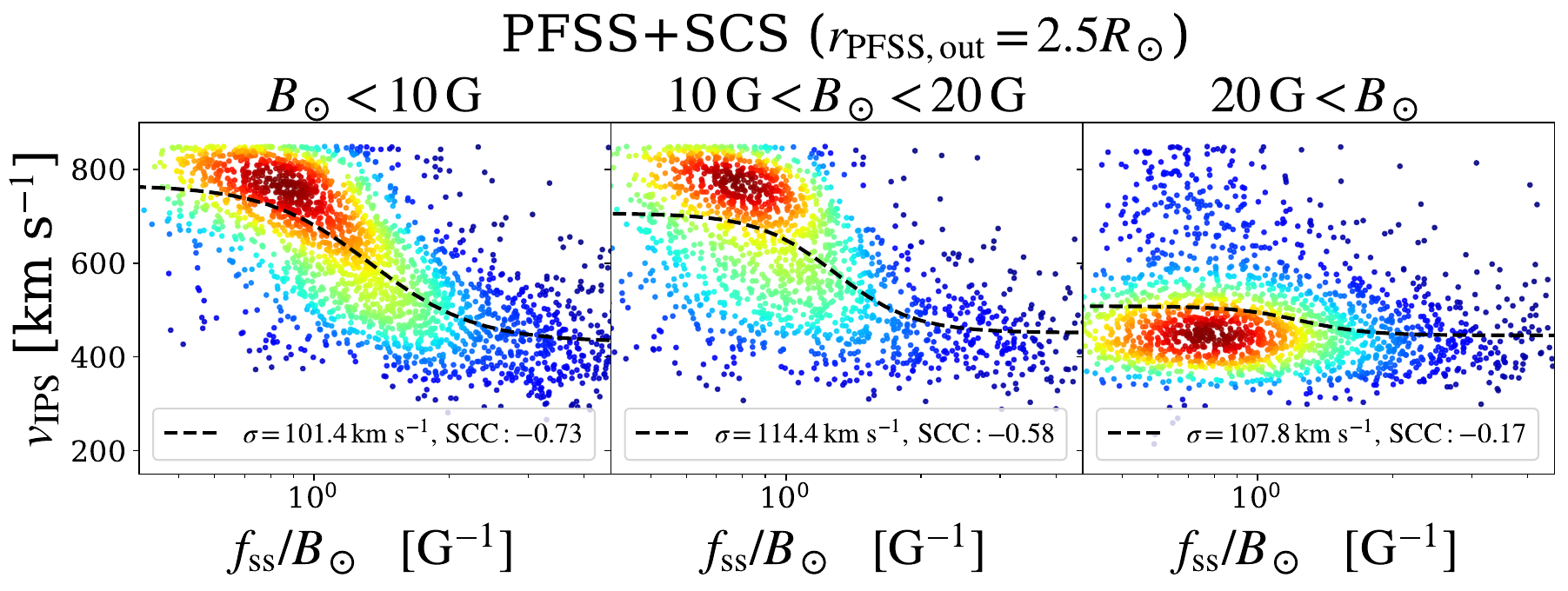}
  \caption{The same as the upper row of Figure \ref{fig: foot point B} but $r_{\rm PFSS,out}$ is fixed at $2.5R_\odot$.
  }
  \label{fig:r_PFSSout2.5}
  \end{center}
\end{figure*}

In the main text, we determine $r_{\rm PFSS,out}$ and $r_{\rm SCS,in}$ based on the open magnetic flux calculated from the OMNI database, following \citet{Shoda_2025_ApJ}. However, we find that essentially the same results are obtained when adopting the commonly used value $r_{\rm PFSS,out} = 2.5R_\odot$. The corresponding results are shown in Figure \ref{fig:r_PFSSout2.5}. As is evident from this figure, as well as from the upper row of Figure \ref{fig: foot point B}, the results of our analysis depend only weakly-both quantitatively and qualitatively-on the method used to determine $r_{\rm PFSS,out}$.

However, this result must be interpreted with the caution in light of the uncertainties in the IPS measurements. \citet{Tokumaru_2021_ApJ} estimated that the uncertainty in the IPS-derived wind speeds prior to 2009 is on the order of $100$ km s$^{-1}$, based on comparisons with in-situ observations such as Ulysses and OMNI. This level of uncertainty is comparable to the RMSE obtained with the PFSS+SCS model in this study. Therefore, the apparent lack of strong dependence on the source-surface height may, in part, reflect the fact that such effects are obscured by observational uncertainties. To resolve variations smaller than $\sim 100$ km s$^{-1}$, higher-resolution and higher-accuracy observational data will be required.

\section{Uncertainty of the threshold of $B_\odot$} \label{app:Bodot}

\begin{figure*}[!t]
  \begin{center}
  \plotone{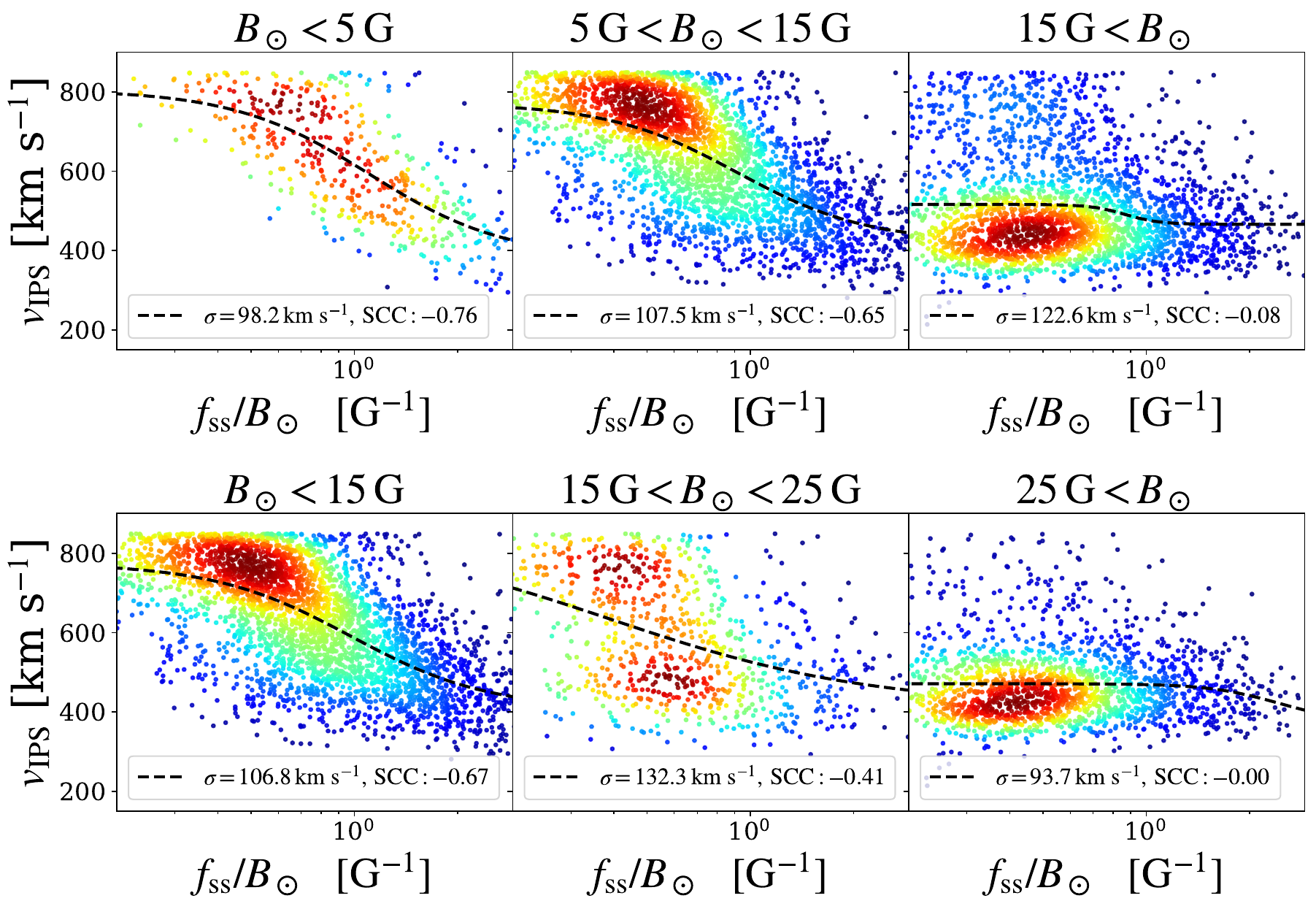}
  \caption{The same as the upper row of Figure \ref{fig: foot point B} but the thresholds of $B_\odot$ are $5$ G and $15$ G (the upper row) and $15$ G and $25$ G (the upper row).
  }
  \label{fig:Bodot_thr}
  \end{center}
\end{figure*}

In Figure \ref{fig: foot point B} of the main text, the analysis was performed using threshold values of $10$ G and $20$ G to classify $B_\odot$, although the rationale for these choices was not explicitly discussed. In this section, we examine the dependence on these thresholds in more detail. Specifically, we consider two alternative cases: thresholds of $5$ G and $15$ G, and thresholds of $15$ G and $25$ G. The corresponding results are shown in the upper and lower rows of Figure \ref{fig:Bodot_thr}, respectively.

Qualitatively, in both cases, the contrasting dependence of the solar wind speed on $f_{\rm SS}/B_\odot$ between weak and strong footpoint magnetic field regions, as described in the main text, is consistently reproduced. This indicates that the specific choice of threshold values, such as $10$ G and $20$ G, is not essential to the main conclusions.

On the other hand, from a quantitative perspective, as the threshold defining the weak footpoint magnetic field increases, the RMSE $\sigma$ systematically becomes larger and the absolute value of the SCC decreases, indicating a weakening of the correlation. This suggests that, for applications to solar wind speed prediction, further exploration and optimization of these threshold values will be necessary.

\bibliography{reference}{}

@ARTICLE{Alazraki_1971_AandA,
       author = {{Alazraki}, G. and {Couturier}, P.},
        title = "{Solar Wind Accejeration Caused by the Gradient of Alfven Wave Pressure}",
      journal = {\aap},
         year = 1971,
        month = aug,
       volume = {13},
        pages = {380},
       adsurl = {https://ui.adsabs.harvard.edu/abs/1971A&A....13..380A}
}

@ARTICLE{Altschuler_1969_SoPh,
       author = {{Altschuler}, Martin D. and {Newkirk}, Jr., Gordon},
        title = "{Magnetic Fields and the Structure of the Solar Corona. I: Methods of Calculating Coronal Fields}",
      journal = {\solphys},
         year = 1969,
        month = sep,
       volume = {9},
       number = {1},
        pages = {131-149},
          doi = {10.1007/BF00145734},
       adsurl = {https://ui.adsabs.harvard.edu/abs/1969SoPh....9..131A}
}

@ARTICLE{Arge_2000_JGR,
       author = {{Arge}, C.~N. and {Pizzo}, V.~J.},
        title = "{Improvement in the prediction of solar wind conditions using near-real time solar magnetic field updates}",
      journal = {\jgr},
         year = 2000,
        month = may,
       volume = {105},
       number = {A5},
        pages = {10465-10480},
          doi = {10.1029/1999JA000262},
       adsurl = {https://ui.adsabs.harvard.edu/abs/2000JGR...10510465A}
}

@INPROCEEDINGS{Arge_2003_AIP,
       author = {{Arge}, Charles N. and {Odstrcil}, Dusan and {Pizzo}, Victor J. and {Mayer}, Leslie R.},
        title = "{Improved Method for Specifying Solar Wind Speed Near the Sun}",
    booktitle = {Solar Wind Ten},
         year = 2003,
       editor = {{Velli}, Marco and {Bruno}, Roberto and {Malara}, Francesco and {Bucci}, B.},
       series = {American Institute of Physics Conference Series},
       volume = {679},
        month = sep,
    publisher = {AIP},
        pages = {190-193},
          doi = {10.1063/1.1618574},
       adsurl = {https://ui.adsabs.harvard.edu/abs/2003AIPC..679..190A}
}

@INPROCEEDINGS{Arge_2010_AIPCS,
       author = {{Arge}, C. Nick and {Henney}, Carl J. and {Koller}, Josef and {Compeau}, C. Rich and {Young}, Shawn and {MacKenzie}, David and {Fay}, Alex and {Harvey}, John W.},
        title = "{Air Force Data Assimilative Photospheric Flux Transport (ADAPT) Model}",
    booktitle = {Twelfth International Solar Wind Conference},
         year = 2010,
       editor = {{Maksimovic}, M. and {Issautier}, K. and {Meyer-Vernet}, N. and {Moncuquet}, M. and {Pantellini}, F.},
       series = {American Institute of Physics Conference Series},
       volume = {1216},
        month = mar,
    publisher = {AIP},
        pages = {343-346},
          doi = {10.1063/1.3395870},
       adsurl = {https://ui.adsabs.harvard.edu/abs/2010AIPC.1216..343A}
}

@INPROCEEDINGS{Arge_2013_AIPC,
       author = {{Arge}, C. Nick and {Henney}, Carl J. and {Hernandez}, Irene Gonzalez and {Toussaint}, W. Alex and {Koller}, Josef and {Godinez}, Humberto C.},
        title = "{Modeling the corona and solar wind using ADAPT maps that include far-side observations}",
    booktitle = {Solar Wind 13},
         year = 2013,
       editor = {{Zank}, Gary P. and {Borovsky}, Joe and {Bruno}, Roberto and {Cirtain}, Jonathan and {Cranmer}, Steve and {Elliott}, Heather and {Giacalone}, Joe and {Gonzalez}, Walter and {Li}, Gang and {Marsch}, Eckart and {Moebius}, Ebehard and {Pogorelov}, Nick and {Spann}, Jim and {Verkhoglyadova}, Olga},
       series = {American Institute of Physics Conference Series},
       volume = {1539},
        month = jun,
    publisher = {AIP},
        pages = {11-14},
          doi = {10.1063/1.4810977},
       adsurl = {https://ui.adsabs.harvard.edu/abs/2013AIPC.1539...11A}
}

@ARTICLE{Asai_1995_JGG,
       author = {{Asai}, Kikuo and {Ishida}, Yoshio and {Kojma}, Masayoshi and {Maruyama}, Kazuo and {Misawa}, Hiroaki and {Yoshimi}, Naohiko},
        title = "{Multi-Station System for Solar Wind Observations Using the Interplanetary Scintillation Method.}",
      journal = {Journal of Geomagnetism and Geoelectricity},
         year = 1995,
        month = jan,
       volume = {47},
       number = {11},
        pages = {1107-1112},
          doi = {10.5636/jgg.47.1107},
       adsurl = {https://ui.adsabs.harvard.edu/abs/1995JGG....47.1107A}
}

@ARTICLE{Bale_2023_Natur,
       author = {{Bale}, S.~D. and {Drake}, J.~F. and {McManus}, M.~D. and {Desai}, M.~I. and {Badman}, S.~T. and {Larson}, D.~E. and {Swisdak}, M. and {Horbury}, T.~S. and {Raouafi}, N.~E. and {Phan}, T. and {Velli}, M. and {McComas}, D.~J. and {Cohen}, C.~M.~S. and {Mitchell}, D. and {Panasenco}, O. and {Kasper}, J.~C.},
        title = "{Interchange reconnection as the source of the fast solar wind within coronal holes}",
      journal = {\nat},
         year = 2023,
        month = jun,
       volume = {618},
       number = {7964},
        pages = {252-256},
          doi = {10.1038/s41586-023-05955-3},
archivePrefix = {arXiv},
       eprint = {2208.07932},
 primaryClass = {astro-ph.SR},
       adsurl = {https://ui.adsabs.harvard.edu/abs/2023Natur.618..252B}
}

@ARTICLE{Chen_2025_AandA,
       author = {{Chen}, Yajie and {Peter}, Hardi and {Przybylski}, Damien and {Iijima}, Haruhisa and {Chitta}, Lakshmi Pradeep},
        title = "{Magnetic reconnection sustains the mass budget of the solar wind}",
      journal = {\aap},
         year = 2025,
        month = oct,
       volume = {702},
          eid = {L4},
        pages = {L4},
          doi = {10.1051/0004-6361/202556696},
archivePrefix = {arXiv},
       eprint = {2509.11692},
 primaryClass = {astro-ph.SR},
       adsurl = {https://ui.adsabs.harvard.edu/abs/2025A&A...702L...4C}
}

@ARTICLE{Cranmer_2002_SSR,
       author = {{Cranmer}, Steven R.},
        title = "{Coronal Holes and the High-Speed Solar Wind}",
      journal = {\ssr},
         year = 2002,
        month = aug,
       volume = {101},
       number = {3},
        pages = {229-294},
          doi = {10.1023/A:1020840004535},
       adsurl = {https://ui.adsabs.harvard.edu/abs/2002SSRv..101..229C}
}

@ARTICLE{DePontieu_2007_Sci,
       author = {{De Pontieu}, B. and {McIntosh}, S.~W. and {Carlsson}, M. and {Hansteen}, V.~H. and {Tarbell}, T.~D. and {Schrijver}, C.~J. and {Title}, A.~M. and {Shine}, R.~A. and {Tsuneta}, S. and {Katsukawa}, Y. and {Ichimoto}, K. and {Suematsu}, Y. and {Shimizu}, T. and {Nagata}, S.},
        title = "{Chromospheric Alfv{\'e}nic Waves Strong Enough to Power the Solar Wind}",
      journal = {Science},
         year = 2007,
        month = dec,
       volume = {318},
       number = {5856},
        pages = {1574},
          doi = {10.1126/science.1151747},
       adsurl = {https://ui.adsabs.harvard.edu/abs/2007Sci...318.1574D}
}

@INPROCEEDINGS{Deforest_2022_aero,
       author = {{Deforest}, Craig and {Killough}, Ronnie and {Gibson}, Sarah and {Henry}, Alan and {Case}, Traci and {Beasley}, Matthew and {Laurent}, Glenn and {Colaninno}, Robin and {Waltham}, Nick and {Punch Science Team}},
        title = "{Polarimeter to UNify the Corona and Heliosphere (PUNCH): Science, Status, and Path to Flight}",
    booktitle = {2022 IEEE Aerospace Conference},
         year = 2022,
        month = aug,
          eid = {1},
        pages = {1-11},
          doi = {10.1109/AERO53065.2022.9843340},
       adsurl = {https://ui.adsabs.harvard.edu/abs/2022aero.confE...1D}
}

@ARTICLE{Fisk_1999_JGR,
       author = {{Fisk}, L.~A. and {Schwadron}, N.~A. and {Zurbuchen}, T.~H.},
        title = "{Acceleration of the fast solar wind by the emergence of new magnetic flux}",
      journal = {\jgr},
         year = 1999,
        month = sep,
       volume = {104},
       number = {A9},
        pages = {19765-19772},
          doi = {10.1029/1999JA900256},
       adsurl = {https://ui.adsabs.harvard.edu/abs/1999JGR...10419765F}
}

@ARTICLE{Fisk_2003_JGR,
       author = {{Fisk}, L.~A.},
        title = "{Acceleration of the solar wind as a result of the reconnection of open magnetic flux with coronal loops}",
      journal = {Journal of Geophysical Research (Space Physics)},
         year = 2003,
        month = apr,
       volume = {108},
       number = {A4},
          eid = {1157},
        pages = {1157},
          doi = {10.1029/2002JA009284},
       adsurl = {https://ui.adsabs.harvard.edu/abs/2003JGRA..108.1157F}
}

@ARTICLE{Fujiki_2015_SoPh,
       author = {{Fujiki}, Ken'ichi and {Tokumaru}, Munetoshi and {Iju}, Tomoya and {Hakamada}, Kazuyuki and {Kojima}, Masayoshi},
        title = "{Relationship Between Solar-Wind Speed and Coronal Magnetic-Field Properties}",
      journal = {\solphys},
         year = 2015,
        month = sep,
       volume = {290},
       number = {9},
        pages = {2491-2505},
          doi = {10.1007/s11207-015-0742-8},
archivePrefix = {arXiv},
       eprint = {1507.03301},
 primaryClass = {astro-ph.SR},
       adsurl = {https://ui.adsabs.harvard.edu/abs/2015SoPh..290.2491F}
}

@ARTICLE{Hansteen_2012_SSR,
       author = {{Hansteen}, Viggo H. and {Velli}, Marco},
        title = "{Solar Wind Models from the Chromosphere to 1 AU}",
      journal = {\ssr},
         year = 2012,
        month = nov,
       volume = {172},
       number = {1-4},
        pages = {89-121},
          doi = {10.1007/s11214-012-9887-z},
       adsurl = {https://ui.adsabs.harvard.edu/abs/2012SSRv..172...89H}
}

@INPROCEEDINGS{Henney_2006_ASPC,
       author = {{Henney}, C.~J. and {Keller}, C.~U. and {Harvey}, J.~W.},
        title = "{SOLIS-VSM Solar Vector Magnetograms}",
    booktitle = {Solar Polarization 4},
         year = 2006,
       editor = {{Casini}, R. and {Lites}, B.~W.},
       series = {Astronomical Society of the Pacific Conference Series},
       volume = {358},
        month = dec,
        pages = {92},
          doi = {10.48550/arXiv.astro-ph/0612584},
archivePrefix = {arXiv},
       eprint = {astro-ph/0612584},
 primaryClass = {astro-ph},
       adsurl = {https://ui.adsabs.harvard.edu/abs/2006ASPC..358...92H}
}

@ARTICLE{Hickmann_2015_SoPh,
       author = {{Hickmann}, Kyle S. and {Godinez}, Humberto C. and {Henney}, Carl J. and {Arge}, C. Nick},
        title = "{Data Assimilation in the ADAPT Photospheric Flux Transport Model}",
      journal = {\solphys},
         year = 2015,
        month = apr,
       volume = {290},
       number = {4},
        pages = {1105-1118},
          doi = {10.1007/s11207-015-0666-3},
archivePrefix = {arXiv},
       eprint = {1410.6185},
 primaryClass = {math-ph},
       adsurl = {https://ui.adsabs.harvard.edu/abs/2015SoPh..290.1105H}
}

@ARTICLE{Hoeksema_1983_JGR,
       author = {{Hoeksema}, J.~T. and {Wilcox}, J.~M. and {Scherrer}, P.~H.},
        title = "{The structure of the heliospheric current sheet: 1978-1982}",
      journal = {\jgr},
         year = 1983,
        month = dec,
       volume = {88},
       number = {A12},
        pages = {9910-9918},
          doi = {10.1029/JA088iA12p09910},
       adsurl = {https://ui.adsabs.harvard.edu/abs/1983JGR....88.9910H}
}

@ARTICLE{Hollweg_1973_JGR,
       author = {{Hollweg}, Joseph V.},
        title = "{Alfv{\'e}n waves in the solar wind: Wave pressure, poynting flux, and angular momentum}",
      journal = {\jgr},
         year = 1973,
        month = jan,
       volume = {78},
       number = {19},
        pages = {3643},
          doi = {10.1029/JA078i019p03643},
       adsurl = {https://ui.adsabs.harvard.edu/abs/1973JGR....78.3643H}
}

@ARTICLE{Huang_2024_ApJL,
       author = {{Huang}, Zhenguang and {T{\'o}th}, G{\'a}bor and {Huang}, Jia and {Sachdeva}, Nishtha and {van der Holst}, Bart and {Manchester}, Ward B.},
        title = "{Adjusting the Potential Field Source Surface Height Based on Magnetohydrodynamic Simulations}",
      journal = {\apjl},
         year = 2024,
        month = apr,
       volume = {965},
       number = {1},
          eid = {L1},
        pages = {L1},
          doi = {10.3847/2041-8213/ad3547},
archivePrefix = {arXiv},
       eprint = {2403.01712},
 primaryClass = {astro-ph.SR},
       adsurl = {https://ui.adsabs.harvard.edu/abs/2024ApJ...965L...1H}
}

@ARTICLE{Iijima_2023_ApJ,
       author = {{Iijima}, Haruhisa and {Matsumoto}, Takuma and {Hotta}, Hideyuki and {Imada}, Shinsuke},
        title = "{A Comprehensive Simulation of Solar Wind Formation from the Solar Interior: Significant Cross-field Energy Transport by Interchange Reconnection near the Sun}",
      journal = {\apjl},
         year = 2023,
        month = jul,
       volume = {951},
       number = {2},
          eid = {L47},
        pages = {L47},
          doi = {10.3847/2041-8213/acdde0},
archivePrefix = {arXiv},
       eprint = {2306.17324},
 primaryClass = {astro-ph.SR},
       adsurl = {https://ui.adsabs.harvard.edu/abs/2023ApJ...951L..47I}
}

@ARTICLE{Jacques_1977_ApJ,
       author = {{Jacques}, S.~A.},
        title = "{Momentum and energy transport by waves in the solar atmosphere and solar wind.}",
      journal = {\apj},
         year = 1977,
        month = aug,
       volume = {215},
        pages = {942-951},
          doi = {10.1086/155430},
       adsurl = {https://ui.adsabs.harvard.edu/abs/1977ApJ...215..942J}
}

@ARTICLE{Jones_1992_SoPh,
       author = {{Jones}, Harrison P. and {Duvall}, Jr., Thomas L. and {Harvey}, John W. and {Mahaffey}, Charles T. and {Schwitters}, Jan D. and {Simmons}, Jorge E.},
        title = "{The NASA/NSO Spectromagnetograph}",
      journal = {\solphys},
         year = 1992,
        month = jun,
       volume = {139},
       number = {2},
        pages = {211-232},
          doi = {10.1007/BF00159149},
       adsurl = {https://ui.adsabs.harvard.edu/abs/1992SoPh..139..211J}
}

@INPROCEEDINGS{Keller_2003_ASPC,
       author = {{Keller}, C.~U. and {Harvey}, J.~W. and {Solis Team}},
        title = "{The SOLIS Vector-Spectromagnetograph}",
    booktitle = {Solar Polarization},
         year = 2003,
       editor = {{Trujillo-Bueno}, Javier and {Sanchez Almeida}, Jorge},
       series = {Astronomical Society of the Pacific Conference Series},
       volume = {307},
        month = jan,
        pages = {13},
       adsurl = {https://ui.adsabs.harvard.edu/abs/2003ASPC..307...13K}
}

@ARTICLE{Knizhnik_2024_FrASS,
       author = {{Knizhnik}, Kalman J.},
        title = "{The Schatten current sheet}",
      journal = {Frontiers in Astronomy and Space Sciences},
         year = 2024,
        month = oct,
       volume = {11},
          eid = {1476498},
        pages = {1476498},
          doi = {10.3389/fspas.2024.1476498},
       adsurl = {https://ui.adsabs.harvard.edu/abs/2024FrASS..1176498K}
}

@ARTICLE{Kojima_1990_SSR,
       author = {{Kojima}, Masayoshi and {Kakinuma}, Takakiyo},
        title = "{Solar Cycle Dependence of Global Distribution of Solar Wind Speed}",
      journal = {\ssr},
         year = 1990,
        month = aug,
       volume = {53},
       number = {3-4},
        pages = {173-222},
          doi = {10.1007/BF00212754},
       adsurl = {https://ui.adsabs.harvard.edu/abs/1990SSRv...53..173K}
}

@ARTICLE{Magyar_2017_NatSR,
       author = {{Magyar}, Norbert and {Van Doorsselaere}, Tom and {Goossens}, Marcel},
        title = "{Generalized phase mixing: Turbulence-like behaviour from unidirectionally propagating MHD waves}",
      journal = {Scientific Reports},
         year = 2017,
        month = nov,
       volume = {7},
          eid = {14820},
        pages = {14820},
          doi = {10.1038/s41598-017-13660-1},
archivePrefix = {arXiv},
       eprint = {1702.02346},
 primaryClass = {astro-ph.SR},
       adsurl = {https://ui.adsabs.harvard.edu/abs/2017NatSR...714820M}
}

@ARTICLE{McComas_2008_GRL,
       author = {{McComas}, D.~J. and {Ebert}, R.~W. and {Elliott}, H.~A. and {Goldstein}, B.~E. and {Gosling}, J.~T. and {Schwadron}, N.~A. and {Skoug}, R.~M.},
        title = "{Weaker solar wind from the polar coronal holes and the whole Sun}",
      journal = {\grl},
         year = 2008,
        month = sep,
       volume = {35},
       number = {18},
          eid = {L18103},
        pages = {L18103},
          doi = {10.1029/2008GL034896},
       adsurl = {https://ui.adsabs.harvard.edu/abs/2008GeoRL..3518103M}
}

@ARTICLE{McGregor_2008_JGR,
       author = {{McGregor}, S.~L. and {Hughes}, W.~J. and {Arge}, C.~N. and {Owens}, M.~J.},
        title = "{Analysis of the magnetic field discontinuity at the potential field source surface and Schatten Current Sheet interface in the Wang-Sheeley-Arge model}",
      journal = {Journal of Geophysical Research (Space Physics)},
         year = 2008,
        month = aug,
       volume = {113},
       number = {A8},
          eid = {A08112},
        pages = {A08112},
          doi = {10.1029/2007JA012330},
       adsurl = {https://ui.adsabs.harvard.edu/abs/2008JGRA..113.8112M}
}

@ARTICLE{McIntosh_2011_Natur,
       author = {{McIntosh}, Scott W. and {de Pontieu}, Bart and {Carlsson}, Mats and {Hansteen}, Viggo and {Boerner}, Paul and {Goossens}, Marcel},
        title = "{Alfv{\'e}nic waves with sufficient energy to power the quiet solar corona and fast solar wind}",
      journal = {\nat},
         year = 2011,
        month = jul,
       volume = {475},
       number = {7357},
        pages = {477-480},
          doi = {10.1038/nature10235},
       adsurl = {https://ui.adsabs.harvard.edu/abs/2011Natur.475..477M}
}

@ARTICLE{Meadors_2020_SW,
       author = {{Meadors}, Grant David and {Jones}, Shaela I. and {Hickmann}, Kyle S. and {Arge}, Charles N. and {Godinez-Vasquez}, Humberto C. and {Henney}, Carl J.},
        title = "{Data Assimilative Optimization of WSA Source Surface and Interface Radii using Particle Filtering}",
      journal = {Space Weather},
         year = 2020,
        month = may,
       volume = {18},
       number = {5},
          eid = {e02464},
        pages = {e02464},
          doi = {10.1029/2020SW00246410.1002/essoar.10502019.1},
       adsurl = {https://ui.adsabs.harvard.edu/abs/2020SpWea..1802464M}
}

@ARTICLE{Morton_2025_ApJ,
       author = {{Morton}, R.~J. and {Weberg}, M.~J. and {Balodhi}, N. and {McLaughlin}, J.~A.},
        title = "{Estimating the Poynting Flux of Alfv{\'e}nic Waves in Polar Coronal Holes across Solar Cycle 24}",
      journal = {\apj},
         year = 2025,
        month = may,
       volume = {985},
       number = {1},
          eid = {13},
        pages = {13},
          doi = {10.3847/1538-4357/adc568},
archivePrefix = {arXiv},
       eprint = {2501.13673},
 primaryClass = {astro-ph.SR},
       adsurl = {https://ui.adsabs.harvard.edu/abs/2025ApJ...985...13M}
}

@ARTICLE{Muller_2020_A&A,
       author = {{M{\"u}ller}, D. and {St. Cyr}, O.~C. and {Zouganelis}, I. and {Gilbert}, H.~R. and {Marsden}, R. and {Nieves-Chinchilla}, T. and {Antonucci}, E. and {Auch{\`e}re}, F. and {Berghmans}, D. and {Horbury}, T.~S. and {Howard}, R.~A. and {Krucker}, S. and {Maksimovic}, M. and {Owen}, C.~J. and {Rochus}, P. and {Rodriguez-Pacheco}, J. and {Romoli}, M. and {Solanki}, S.~K. and {Bruno}, R. and {Carlsson}, M. and {Fludra}, A. and {Harra}, L. and {Hassler}, D.~M. and {Livi}, S. and {Louarn}, P. and {Peter}, H. and {Sch{\"u}hle}, U. and {Teriaca}, L. and {del Toro Iniesta}, J.~C. and {Wimmer-Schweingruber}, R.~F. and {Marsch}, E. and {Velli}, M. and {De Groof}, A. and {Walsh}, A. and {Williams}, D.},
        title = "{The Solar Orbiter mission. Science overview}",
      journal = {\aap},
         year = 2020,
        month = oct,
       volume = {642},
          eid = {A1},
        pages = {A1},
          doi = {10.1051/0004-6361/202038467},
archivePrefix = {arXiv},
       eprint = {2009.00861},
 primaryClass = {astro-ph.SR},
       adsurl = {https://ui.adsabs.harvard.edu/abs/2020A&A...642A...1M}
}

@ARTICLE{Owens_2008_SW,
       author = {{Owens}, M.~J. and {Spence}, H.~E. and {McGregor}, S. and {Hughes}, W.~J. and {Quinn}, J.~M. and {Arge}, C.~N. and {Riley}, P. and {Linker}, J. and {Odstrcil}, D.},
        title = "{Metrics for solar wind prediction models: Comparison of empirical, hybrid, and physics-based schemes with 8 years of L1 observations}",
      journal = {Space Weather},
         year = 2008,
        month = aug,
       volume = {6},
       number = {8},
          eid = {S08001},
        pages = {S08001},
          doi = {10.1029/2007SW000380},
       adsurl = {https://ui.adsabs.harvard.edu/abs/2008SpWea...6.8001O}
}

@ARTICLE{Owens_2020_SoPh,
       author = {{Owens}, Mathew and {Lockwood}, Mike and {Macneil}, Allan and {Stansby}, David},
        title = "{Signatures of Coronal Loop Opening via Interchange Reconnection in the Slow Solar Wind at 1 AU}",
      journal = {\solphys},
         year = 2020,
        month = mar,
       volume = {295},
       number = {3},
          eid = {37},
        pages = {37},
          doi = {10.1007/s11207-020-01601-7},
       adsurl = {https://ui.adsabs.harvard.edu/abs/2020SoPh..295...37O}
}

@ARTICLE{Parker_1958_ApJ,
       author = {{Parker}, E.~N.},
        title = "{Dynamics of the Interplanetary Gas and Magnetic Fields.}",
      journal = {\apj},
         year = 1958,
        month = nov,
       volume = {128},
        pages = {664},
          doi = {10.1086/146579},
       adsurl = {https://ui.adsabs.harvard.edu/abs/1958ApJ...128..664P}
}

@ARTICLE{Rappazzo_2012_ApJ,
       author = {{Rappazzo}, A.~F. and {Matthaeus}, W.~H. and {Ruffolo}, D. and {Servidio}, S. and {Velli}, M.},
        title = "{Interchange Reconnection in a Turbulent Corona}",
      journal = {\apjl},
         year = 2012,
        month = oct,
       volume = {758},
       number = {1},
          eid = {L14},
        pages = {L14},
          doi = {10.1088/2041-8205/758/1/L14},
archivePrefix = {arXiv},
       eprint = {1209.1388},
 primaryClass = {astro-ph.SR},
       adsurl = {https://ui.adsabs.harvard.edu/abs/2012ApJ...758L..14R}
}

@ARTICLE{Rice_2021_ApJ,
       author = {{Rice}, Oliver E.~K. and {Yeates}, Anthony R.},
        title = "{Global Coronal Equilibria with Solar Wind Outflow}",
      journal = {\apj},
         year = 2021,
        month = dec,
       volume = {923},
       number = {1},
          eid = {57},
        pages = {57},
          doi = {10.3847/1538-4357/ac2c71},
archivePrefix = {arXiv},
       eprint = {2110.01319},
 primaryClass = {astro-ph.SR},
       adsurl = {https://ui.adsabs.harvard.edu/abs/2021ApJ...923...57R}
}

@ARTICLE{Rice_2026_arXiv,
       author = {{Rice}, Oliver and {Yeates}, Anthony},
        title = "{Global Coronal Equilibria with Solar Wind Outflow II -- Optimizing the Outflow Model}",
      journal = {arXiv e-prints},
         year = 2026,
        month = mar,
          eid = {arXiv:2603.22159},
        pages = {arXiv:2603.22159},
          doi = {10.48550/arXiv.2603.22159},
archivePrefix = {arXiv},
       eprint = {2603.22159},
 primaryClass = {astro-ph.SR},
       adsurl = {https://ui.adsabs.harvard.edu/abs/2026arXiv260322159R}
}

@ARTICLE{Rivera_2024_Science,
       author = {{Rivera}, Yeimy J. and {Badman}, Samuel T. and {Stevens}, Michael L. and {Verniero}, Jaye L. and {Stawarz}, Julia E. and {Shi}, Chen and {Raines}, Jim M. and {Paulson}, Kristoff W. and {Owen}, Christopher J. and {Niembro}, Tatiana and {Louarn}, Philippe and {Livi}, Stefano A. and {Lepri}, Susan T. and {Kasper}, Justin C. and {Horbury}, Timothy S. and {Halekas}, Jasper S. and {Dewey}, Ryan M. and {De Marco}, Rossana and {Bale}, Stuart D.},
        title = "{In situ observations of large-amplitude Alfv{\'e}n waves heating and accelerating the solar wind}",
      journal = {Science},
         year = 2024,
        month = aug,
       volume = {385},
       number = {6712},
        pages = {962-966},
          doi = {10.1126/science.adk6953},
archivePrefix = {arXiv},
       eprint = {2409.00267},
 primaryClass = {astro-ph.SR},
       adsurl = {https://ui.adsabs.harvard.edu/abs/2024Sci...385..962R}
}

@ARTICLE{Rivera_2025_ApJ,
       author = {{Rivera}, Yeimy J. and {Badman}, Samuel T. and {Verniero}, J.~L. and {Varesano}, Tania and {Stevens}, Michael L. and {Stawarz}, Julia E. and {Reeves}, Katharine K. and {Raines}, Jim M. and {Raymond}, John C. and {Owen}, Christopher J. and {Livi}, Stefano A. and {Lepri}, Susan T. and {Landi}, Enrico and {Halekas}, Jasper. S. and {Ervin}, Tamar and {Dewey}, Ryan M. and {De Marco}, Rossana and {D'Amicis}, Raffaella and {Dakeyo}, Jean-Baptiste and {Bale}, Stuart D. and {Alterman}, B.~L.},
        title = "{Differentiating the Acceleration Mechanisms in the Slow and Alfv{\'e}nic Slow Solar Wind}",
      journal = {\apj},
         year = 2025,
        month = feb,
       volume = {980},
       number = {1},
          eid = {70},
        pages = {70},
          doi = {10.3847/1538-4357/ada699},
archivePrefix = {arXiv},
       eprint = {2501.02163},
 primaryClass = {astro-ph.SR},
       adsurl = {https://ui.adsabs.harvard.edu/abs/2025ApJ...980...70R}
}

@ARTICLE{Reville_2020_ApJS,
       author = {{R{\'e}ville}, Victor and {Velli}, Marco and {Panasenco}, Olga and {Tenerani}, Anna and {Shi}, Chen and {Badman}, Samuel T. and {Bale}, Stuart D. and {Kasper}, J.~C. and {Stevens}, Michael L. and {Korreck}, Kelly E. and {Bonnell}, J.~W. and {Case}, Anthony W. and {de Wit}, Thierry Dudok and {Goetz}, Keith and {Harvey}, Peter R. and {Larson}, Davin E. and {Livi}, Roberto and {Malaspina}, David M. and {MacDowall}, Robert J. and {Pulupa}, Marc and {Whittlesey}, Phyllis L.},
        title = "{The Role of Alfv{\'e}n Wave Dynamics on the Large-scale Properties of the Solar Wind: Comparing an MHD Simulation with Parker Solar Probe E1 Data}",
      journal = {\apjs},
         year = 2020,
        month = feb,
       volume = {246},
       number = {2},
          eid = {24},
        pages = {24},
          doi = {10.3847/1538-4365/ab4fef},
archivePrefix = {arXiv},
       eprint = {1912.03777},
 primaryClass = {astro-ph.SR},
       adsurl = {https://ui.adsabs.harvard.edu/abs/2020ApJS..246...24R}
}

@ARTICLE{Riley_2001_JGR,
       author = {{Riley}, Pete and {Linker}, J.~A. and {Miki{\'c}}, Z.},
        title = "{An empirically-driven global MHD model of the solar corona and inner heliosphere}",
      journal = {\jgr},
         year = 2001,
        month = aug,
       volume = {106},
       number = {A8},
        pages = {15889-15902},
          doi = {10.1029/2000JA000121},
       adsurl = {https://ui.adsabs.harvard.edu/abs/2001JGR...10615889R}
}

@ARTICLE{Riley_2007_JASTP,
       author = {{Riley}, Pete},
        title = "{Modeling corotating interaction regions: From the Sun to 1 AU}",
      journal = {Journal of Atmospheric and Solar-Terrestrial Physics},
         year = 2007,
        month = feb,
       volume = {69},
       number = {1-2},
        pages = {32-42},
          doi = {10.1016/j.jastp.2006.06.008},
       adsurl = {https://ui.adsabs.harvard.edu/abs/2007JASTP..69...32R}
}

@ARTICLE{Riley_2008_ApJ,
       author = {{Riley}, Pete and {Linker}, J.~A. and {Miki{\'c}}, Z. and {Lionello}, R. and {Ledvina}, S.~A. and {Luhmann}, J.~G.},
        title = "{A Comparison between Global Solar Magnetohydrodynamic and Potential Field Source Surface Model Results}",
      journal = {\apj},
         year = 2006,
        month = dec,
       volume = {653},
       number = {2},
        pages = {1510-1516},
          doi = {10.1086/508565},
       adsurl = {https://ui.adsabs.harvard.edu/abs/2006ApJ...653.1510R}
}

@ARTICLE{Riley_2012_SoPh,
       author = {{Riley}, P. and {Luhmann}, J.~G.},
        title = "{Interplanetary Signatures of Unipolar Streamers and the Origin of the Slow Solar Wind}",
      journal = {\solphys},
         year = 2012,
        month = apr,
       volume = {277},
       number = {2},
        pages = {355-373},
          doi = {10.1007/s11207-011-9909-0},
       adsurl = {https://ui.adsabs.harvard.edu/abs/2012SoPh..277..355R}
}

@ARTICLE{Riley_2015_SW,
       author = {{Riley}, Pete and {Linker}, Jon A. and {Arge}, C. Nick},
        title = "{On the role played by magnetic expansion factor in the prediction of solar wind speed}",
      journal = {Space Weather},
         year = 2015,
        month = mar,
       volume = {13},
       number = {3},
        pages = {154-169},
          doi = {10.1002/2014SW001144},
       adsurl = {https://ui.adsabs.harvard.edu/abs/2015SpWea..13..154R}
}

@ARTICLE{Schatten_1969_SoPh,
       author = {{Schatten}, Kenneth H. and {Wilcox}, John M. and {Ness}, Norman F.},
        title = "{A model of interplanetary and coronal magnetic fields}",
      journal = {\solphys},
         year = 1969,
        month = mar,
       volume = {6},
       number = {3},
        pages = {442-455},
          doi = {10.1007/BF00146478},
       adsurl = {https://ui.adsabs.harvard.edu/abs/1969SoPh....6..442S}
}

@ARTICLE{Schatten_1971_CosEl,
       author = {{Schatten}, K.~H.},
        title = "{Current sheet magnetic model for the solar corona.}",
      journal = {Cosmic Electrodynamics},
         year = 1971,
        month = jan,
       volume = {2},
        pages = {232-245},
       adsurl = {https://ui.adsabs.harvard.edu/abs/1971CosEl...2..232S}
}

@INCOLLECTION{Schatten_1972_NASSP,
       author = {{Schatten}, Kenneth H.},
        title = "{Current Sheet Magnetic Model for the Solar Corona}",
    booktitle = {NASA Special Publication},
         year = 1972,
       editor = {{Sonett}, Charles P. and {Coleman}, Paul Jerome and {Wilcox}, John Marsh},
       volume = {308},
        pages = {44},
       adsurl = {https://ui.adsabs.harvard.edu/abs/1972NASSP.308...44S}
}

@ARTICLE{Shi_2024_ApJ,
       author = {{Shi}, Guanglu and {Feng}, Li and {Ying}, Beili and {Li}, Shuting and {Gan}, Weiqun},
        title = "{Refinement of Global Coronal and Interplanetary Magnetic Field Extrapolations Constrained by Remote-sensing and In Situ Observations at the Solar Minimum}",
      journal = {\apj},
         year = 2024,
        month = aug,
       volume = {970},
       number = {2},
          eid = {131},
        pages = {131},
          doi = {10.3847/1538-4357/ad5200},
archivePrefix = {arXiv},
       eprint = {2405.18665},
 primaryClass = {astro-ph.SR},
       adsurl = {https://ui.adsabs.harvard.edu/abs/2024ApJ...970..131S}
}

@ARTICLE{Shiota_2014_SW,
       author = {{Shiota}, D. and {Kataoka}, R. and {Miyoshi}, Y. and {Hara}, T. and {Tao}, C. and {Masunaga}, K. and {Futaana}, Y. and {Terada}, N.},
        title = "{Inner heliosphere MHD modeling system applicable to space weather forecasting for the other planets}",
      journal = {Space Weather},
         year = 2014,
        month = apr,
       volume = {12},
       number = {4},
        pages = {187-204},
          doi = {10.1002/2013SW000989},
       adsurl = {https://ui.adsabs.harvard.edu/abs/2014SpWea..12..187S}
}

@ARTICLE{Shiota_2016_SW,
       author = {{Shiota}, D. and {Kataoka}, R.},
        title = "{Magnetohydrodynamic simulation of interplanetary propagation of multiple coronal mass ejections with internal magnetic flux rope (SUSANOO-CME)}",
      journal = {Space Weather},
         year = 2016,
        month = feb,
       volume = {14},
       number = {2},
        pages = {56-75},
          doi = {10.1002/2015SW001308},
       adsurl = {https://ui.adsabs.harvard.edu/abs/2016SpWea..14...56S}
}

@ARTICLE{Shoda_2019_ApJ,
       author = {{Shoda}, Munehito and {Suzuki}, Takeru Ken and {Asgari-Targhi}, Mahboubeh and {Yokoyama}, Takaaki},
        title = "{Three-dimensional Simulation of the Fast Solar Wind Driven by Compressible Magnetohydrodynamic Turbulence}",
      journal = {\apjl},
         year = 2019,
        month = jul,
       volume = {880},
       number = {1},
          eid = {L2},
        pages = {L2},
          doi = {10.3847/2041-8213/ab2b45},
archivePrefix = {arXiv},
       eprint = {1905.11685},
 primaryClass = {astro-ph.SR},
       adsurl = {https://ui.adsabs.harvard.edu/abs/2019ApJ...880L...2S}
}

@ARTICLE{Shoda_2023_ApJ,
       author = {{Shoda}, Munehito and {Cranmer}, Steven R. and {Toriumi}, Shin},
        title = "{Formulating Mass-loss Rates for Sun-like Stars: A Hybrid Model Approach}",
      journal = {\apj},
         year = 2023,
        month = nov,
       volume = {957},
       number = {2},
          eid = {71},
        pages = {71},
          doi = {10.3847/1538-4357/acfa72},
archivePrefix = {arXiv},
       eprint = {2309.09399},
 primaryClass = {astro-ph.SR},
       adsurl = {https://ui.adsabs.harvard.edu/abs/2023ApJ...957...71S}
}

@ARTICLE{Shoda_2025_ApJ,
       author = {{Shoda}, Munehito and {Tokoro}, Kyogo and {Shiota}, Daikou and {Imada}, Shinsuke},
        title = "{Empirical Optimization of the Source-surface Height in the Potential Field Source Surface Extrapolation}",
      journal = {\apj},
         year = 2025,
        month = nov,
       volume = {993},
       number = {2},
          eid = {242},
        pages = {242},
          doi = {10.3847/1538-4357/ae10ba},
archivePrefix = {arXiv},
       eprint = {2510.05513},
 primaryClass = {astro-ph.SR},
       adsurl = {https://ui.adsabs.harvard.edu/abs/2025ApJ...993..242S}
}

@misc{SILSO_Sunspot_Number,
        author = {{Clette}, F. and {Lefèvre}, L.},
        title = {SILSO Sunspot Number V2.0},
        doi = {10.24414/qnza-ac80},
        howpublished = {https://doi.org/10.24414/qnza-ac80},
        month = {07},
        year = {2015},
        note = {Published by WDC SILSO - Royal Observatory of Belgium (ROB)}
}

@ARTICLE{Smith_1995_GRL,
       author = {{Smith}, Edward J. and {Balogh}, A.},
        title = "{Ulysses observations of the radial magnetic field}",
      journal = {\grl},
         year = 1995,
        month = dec,
       volume = {22},
       number = {23},
        pages = {3317-3320},
          doi = {10.1029/95GL02826},
       adsurl = {https://ui.adsabs.harvard.edu/abs/1995GeoRL..22.3317S}
}

@ARTICLE{Suzuki_2005_ApJ,
       author = {{Suzuki}, Takeru K. and {Inutsuka}, Shu-ichiro},
        title = "{Making the Corona and the Fast Solar Wind: A Self-consistent Simulation for the Low-Frequency Alfv{\'e}n Waves from the Photosphere to 0.3 AU}",
      journal = {\apjl},
         year = 2005,
        month = oct,
       volume = {632},
       number = {1},
        pages = {L49-L52},
          doi = {10.1086/497536},
archivePrefix = {arXiv},
       eprint = {astro-ph/0506639},
 primaryClass = {astro-ph},
       adsurl = {https://ui.adsabs.harvard.edu/abs/2005ApJ...632L..49S}
}

@ARTICLE{Suzuki_2006_ApJ,
       author = {{Suzuki}, Takeru K.},
        title = "{Forecasting Solar Wind Speeds}",
      journal = {\apjl},
         year = 2006,
        month = mar,
       volume = {640},
       number = {1},
        pages = {L75-L78},
          doi = {10.1086/503102},
archivePrefix = {arXiv},
       eprint = {astro-ph/0602062},
 primaryClass = {astro-ph},
       adsurl = {https://ui.adsabs.harvard.edu/abs/2006ApJ...640L..75S}
}

@ARTICLE{Tokoro_2026_ApJ,
       author = {{Tokoro}, Kyogo and {Shoda}, Munehito and {Imada}, Shinsuke},
        title = "{Proposal of a Novel Physical Parameter Characterizing Solar Wind Speed in a Wave-driven Model}",
      journal = {\apj},
         year = 2026,
        month = feb,
       volume = {997},
       number = {2},
          eid = {351},
        pages = {351},
          doi = {10.3847/1538-4357/ae2fea},
archivePrefix = {arXiv},
       eprint = {2601.21229},
 primaryClass = {astro-ph.SR},
       adsurl = {https://ui.adsabs.harvard.edu/abs/2026ApJ...997..351T}
}

@ARTICLE{Tokumaru_2011_RaSc,
       author = {{Tokumaru}, M. and {Kojima}, M. and {Fujiki}, K. and {Maruyama}, K. and {Maruyama}, Y. and {Ito}, H. and {Iju}, T.},
        title = "{A newly developed UHF radiotelescope for interplanetary scintillation observations: Solar Wind Imaging Facility}",
      journal = {Radio Science},
         year = 2011,
        month = aug,
       volume = {46},
          eid = {RS0F02},
        pages = {RS0F02},
          doi = {10.1029/2011RS004694},
       adsurl = {https://ui.adsabs.harvard.edu/abs/2011RaSc...46.0F02T}
}

@ARTICLE{Tokumaru_2013_PJAB,
       author = {{Tokumaru}, Munetoshi},
        title = "{Three-dimensional exploration of the solar wind using observations of interplanetary scintillation}",
      journal = {Proceedings of the Japan Academy, Series B},
         year = 2013,
        month = jan,
       volume = {89},
       number = {2},
        pages = {67-79},
          doi = {10.2183/pjab.89.67},
       adsurl = {https://ui.adsabs.harvard.edu/abs/2013PJAB...89...67T}
}

@ARTICLE{Tokumaru_2021_ApJ,
       author = {{Tokumaru}, Munetoshi and {Fujiki}, Ken'ichi and {Kojima}, Masayoshi and {Iwai}, Kazumasa},
        title = "{Global Distribution of the Solar Wind Speed Reconstructed from Improved Tomographic Analysis of Interplanetary Scintillation Observations between 1985 and 2019}",
      journal = {\apj},
         year = 2021,
        month = nov,
       volume = {922},
       number = {1},
          eid = {73},
        pages = {73},
          doi = {10.3847/1538-4357/ac1862},
       adsurl = {https://ui.adsabs.harvard.edu/abs/2021ApJ...922...73T}
}

@ARTICLE{Tokumaru_2024_SoPh_DCHB,
       author = {{Tokumaru}, Munetoshi and {Fujiki}, Ken'ichi and {Watanabe}, Haruto},
        title = "{Optimization of Solar-Wind Speed Models Using Interplanetary Scintillation Observations}",
      journal = {\solphys},
         year = 2024,
        month = aug,
       volume = {299},
       number = {8},
          eid = {110},
        pages = {110},
          doi = {10.1007/s11207-024-02356-1},
       adsurl = {https://ui.adsabs.harvard.edu/abs/2024SoPh..299..110T}
}

@ARTICLE{Tokumaru_2024_SoPh_pseudo,
       author = {{Tokumaru}, Munetoshi and {Fujiki}, Ken'ichi},
        title = "{Coronal Magnetic-Field Configuration Associated with Pseudostreamer and Slow Solar Wind}",
      journal = {\solphys},
         year = 2024,
        month = nov,
       volume = {299},
       number = {11},
          eid = {160},
        pages = {160},
          doi = {10.1007/s11207-024-02398-5},
       adsurl = {https://ui.adsabs.harvard.edu/abs/2024SoPh..299..160T}
}

@ARTICLE{Usmanov_2014_ApJ,
       author = {{Usmanov}, Arcadi V. and {Goldstein}, Melvyn L. and {Matthaeus}, William H.},
        title = "{Three-fluid, Three-dimensional Magnetohydrodynamic Solar Wind Model with Eddy Viscosity and Turbulent Resistivity}",
      journal = {\apj},
         year = 2014,
        month = jun,
       volume = {788},
       number = {1},
          eid = {43},
        pages = {43},
          doi = {10.1088/0004-637X/788/1/43},
       adsurl = {https://ui.adsabs.harvard.edu/abs/2014ApJ...788...43U}
}

@ARTICLE{Virtanen_2020_NatMe,
       author = {{Virtanen}, Pauli and {Gommers}, Ralf and {Oliphant}, Travis E. and {Haberland}, Matt and {Reddy}, Tyler and {Cournapeau}, David and {Burovski}, Evgeni and {Peterson}, Pearu and {Weckesser}, Warren and {Bright}, Jonathan and {van der Walt}, St{\'e}fan J. and {Brett}, Matthew and {Wilson}, Joshua and {Millman}, K. Jarrod and {Mayorov}, Nikolay and {Nelson}, Andrew R.~J. and {Jones}, Eric and {Kern}, Robert and {Larson}, Eric and {Carey}, C.~J. and {Polat}, {\.I}lhan and {Feng}, Yu and {Moore}, Eric W. and {VanderPlas}, Jake and {Laxalde}, Denis and {Perktold}, Josef and {Cimrman}, Robert and {Henriksen}, Ian and {Quintero}, E.~A. and {Harris}, Charles R. and {Archibald}, Anne M. and {Ribeiro}, Ant{\^o}nio H. and {Pedregosa}, Fabian and {van Mulbregt}, Paul and {SciPy 1.  0 Contributors}},
        title = "{SciPy 1.0: fundamental algorithms for scientific computing in Python}",
      journal = {Nature Medicine},
         year = 2020,
        month = feb,
       volume = {17},
        pages = {261-272},
          doi = {10.1038/s41592-019-0686-2},
archivePrefix = {arXiv},
       eprint = {1907.10121},
 primaryClass = {cs.MS},
       adsurl = {https://ui.adsabs.harvard.edu/abs/2020NatMe..17..261V}
}

@ARTICLE{Vrsnak_2007_A&A,
       author = {{Vr{\v{s}}nak}, B. and {{\v{Z}}ic}, T.},
        title = "{Transit times of interplanetary coronal mass ejections and the solar wind speed}",
      journal = {\aap},
         year = 2007,
        month = sep,
       volume = {472},
       number = {3},
        pages = {937-943},
          doi = {10.1051/0004-6361:20077499},
       adsurl = {https://ui.adsabs.harvard.edu/abs/2007A&A...472..937V}
}

@ARTICLE{Wang_1990_ApJ,
       author = {{Wang}, Y. -M. and {Sheeley}, Jr., N.~R.},
        title = "{Solar Wind Speed and Coronal Flux-Tube Expansion}",
      journal = {\apj},
         year = 1990,
        month = jun,
       volume = {355},
        pages = {726},
          doi = {10.1086/168805},
       adsurl = {https://ui.adsabs.harvard.edu/abs/1990ApJ...355..726W}
}

@ARTICLE{Wang_2004_ApJ,
       author = {{Wang}, Y.-M. and {Sheeley}, Jr., N.~R.},
        title = "{Footpoint Switching and the Evolution of Coronal Holes}",
      journal = {\apj},
         year = 2004,
        month = sep,
       volume = {612},
       number = {2},
        pages = {1196-1205},
          doi = {10.1086/422711},
       adsurl = {https://ui.adsabs.harvard.edu/abs/2004ApJ...612.1196W}
}

@ARTICLE{Wang_2007_ApJ,
       author = {{Wang}, Y. -M. and {Sheeley}, Jr., N.~R. and {Rich}, N.~B.},
        title = "{Coronal Pseudostreamers}",
      journal = {\apj},
         year = 2007,
        month = apr,
       volume = {658},
       number = {2},
        pages = {1340-1348},
          doi = {10.1086/511416},
       adsurl = {https://ui.adsabs.harvard.edu/abs/2007ApJ...658.1340W}
}

@ARTICLE{Wang_2012_ApJ,
       author = {{Wang}, Y. -M. and {Grappin}, R. and {Robbrecht}, E. and {Sheeley}, Jr., N.~R.},
        title = "{On the Nature of the Solar Wind from Coronal Pseudostreamers}",
      journal = {\apj},
         year = 2012,
        month = apr,
       volume = {749},
       number = {2},
          eid = {182},
        pages = {182},
          doi = {10.1088/0004-637X/749/2/182},
       adsurl = {https://ui.adsabs.harvard.edu/abs/2012ApJ...749..182W}
}

@ARTICLE{Wang_2020_ApJ,
       author = {{Wang}, Y. -M.},
        title = "{Small-scale Flux Emergence, Coronal Hole Heating, and Flux-tube Expansion: A Hybrid Solar Wind Model}",
      journal = {\apj},
         year = 2020,
        month = dec,
       volume = {904},
       number = {2},
          eid = {199},
        pages = {199},
          doi = {10.3847/1538-4357/abbda6},
archivePrefix = {arXiv},
       eprint = {2104.04016},
 primaryClass = {astro-ph.SR},
       adsurl = {https://ui.adsabs.harvard.edu/abs/2020ApJ...904..199W}
}

@ARTICLE{Worden_2000_SoPh,
       author = {{Worden}, John and {Harvey}, John},
        title = "{An Evolving Synoptic Magnetic Flux map and Implications for the Distribution of Photospheric Magnetic Flux}",
      journal = {\solphys},
         year = 2000,
        month = aug,
       volume = {195},
       number = {2},
        pages = {247-268},
          doi = {10.1023/A:1005272502885},
       adsurl = {https://ui.adsabs.harvard.edu/abs/2000SoPh..195..247W}
}
\bibliographystyle{aasjournal}

\end{document}


\setcounter{table}{0}
\renewcommand{\thetable}{S\arabic{table}}

\begin{deluxetable*}{ccc}
\tabletypesize{\scriptsize}
\tablecaption{List of Analyzed Carrington Rotations\label{tab:CR_list}}
\tablehead{
\colhead{Carrington Rotation} & \colhead{Start Date (UT)} & \colhead{$r_{\rm PFSS,out}/R_\odot$} 
}
\startdata
2088 & 2009-09-16 & 1.8 \\
2083 & 2009-05-03 & 1.6 \\
2075 & 2008-09-26 & 1.7 \\
2070 & 2008-05-13 & 1.9 \\
2061 & 2007-09-11 & 2.1 \\
2057 & 2007-05-24 & 1.5 \\
2048 & 2006-09-21 & 1.9 \\
2044 & 2006-06-04 & 1.8 \\
2035 & 2005-10-01 & 1.6 \\
2030 & 2005-05-18 & 2.0 \\
2021 & 2004-09-15 & 2.0 \\
2017 & 2004-05-28 & 1.6 \\
2006 & 2003-08-02 & 1.9 \\
1994 & 2002-09-09 & 1.8 \\
1990 & 2002-05-23 & 1.8 \\
1981 & 2001-09-19 & 1.8 \\
1975 & 2001-04-09 & 1.8 \\
1968 & 2000-09-30 & 2.0 \\
1963 & 2000-05-17 & 1.9 \\
1949 & 1999-05-01 & 1.8 \\
1941 & 1998-09-24 & 1.6 \\
1936 & 1998-05-11 & 2.4 \\
1927 & 1997-09-08 & 1.5 \\
1923 & 1997-05-22 & 2.4 \\
1914 & 1996-09-18 & 2.0 \\
1909 & 1996-05-05 & 1.5 \\
1901 & 1995-09-29 & 1.4 \\
1896 & 1995-05-16 & 1.5 \\
1887 & 1994-09-13 & 1.7 \\
1870 & 1993-06-06 & 1.9 \\
1861 & 1992-10-03 & 1.7 \\
1857 & 1992-06-16 & 1.8 \\
\enddata
\tablecomments{
Start dates are the nominal Carrington rotation start times in UT.
$r_{\mathrm{PFSSout}}$ denotes the outer boundary radius of the PFSS model (in units of~$R_{\odot}$).
}
\end{deluxetable*}